\documentclass[12pt,preprint]{emulateapj}
\usepackage{graphicx}

\newcommand{\pc}{$P_{c}$ }
\newcommand{\mc}{$M_{c} \sin{i}_{c}$ }
\newcommand{\kai}{$\chi^2$ }

\shorttitle{Limits on Stellar Companions}
\shortauthors{Stephen R. Kane et al.}
\slugcomment{Submitted for publication in the Astrophysical Journal}
\begin{document}

\title{Limits on Stellar Companions to Exoplanet Host Stars With
  Eccentric Planets}
\author{
  Stephen R. Kane\altaffilmark{1},
  Steve B. Howell\altaffilmark{2},
  Elliott P. Horch\altaffilmark{3},
  Ying Feng\altaffilmark{4,5},
  Natalie R. Hinkel\altaffilmark{1},
  David R. Ciardi\altaffilmark{6},
  Mark E. Everett\altaffilmark{7},
  Andrew W. Howard\altaffilmark{8},
  Jason T. Wright\altaffilmark{4,5}
}
\email{skane@sfsu.edu}
\altaffiltext{1}{Department of Physics \& Astronomy, San Francisco
  State University, 1600 Holloway Avenue, San Francisco, CA 94132,
  USA}
\altaffiltext{2}{NASA Ames Research Center, Moffett Field, CA 94035,
  USA}
\altaffiltext{3}{Department of Physics, Southern Connecticut State
  University, New Haven, CT 06515, USA}
\altaffiltext{4}{Department of Astronomy and Astrophysics,
  Pennsylvania State University, 525 Davey Laboratory, University
  Park, PA 16802, USA}
\altaffiltext{5}{Center for Exoplanets \& Habitable Worlds,
  Pennsylvania State University, 525 Davey Laboratory, University
  Park, PA 16802, USA}
\altaffiltext{6}{NASA Exoplanet Science Institute, Caltech, MS 100-22,
  770 South Wilson Avenue, Pasadena, CA 91125, USA}
\altaffiltext{7}{National Optical Astronomy Observatory, 950 N. Cherry
  Ave, Tucson, AZ 85719}
\altaffiltext{8}{Institute for Astronomy, University of Hawaii,
  Honolulu, HI 96822, USA}

%%%%%%%%%%%%%%%%%%%%%%%%%%%%%%%%%%%%%%%%%%%%%%%%%%%%%%%%%%%%%%%%%%%%

\begin{abstract}

Though there are now many hundreds of confirmed exoplanets known, the
binarity of exoplanet host stars is not well understood. This is
particularly true of host stars which harbor a giant planet in a
highly eccentric orbit since these are more likely to have had a
dramatic dynamical history which transferred angular momentum to the
planet. Here we present observations of four exoplanet host stars
which utilize the excellent resolving power of the Differential
Speckle Survey Instrument (DSSI) on the Gemini North telescope. Two of
the stars are giants and two are dwarfs. Each star is host to a giant
planet with an orbital eccentricity $> 0.5$ and whose radial velocity
data contain a trend in the residuals to the Keplerian orbit
fit. These observations rule out stellar companions 4--8 magnitudes
fainter than the host star at passbands of 692~nm and 880~nm. The
resolution and field-of-view of the instrument result in exclusion
radii of 0.05--$1.4\arcsec$ which excludes stellar companions within
several AU of the host star in most cases. We further provide new
radial velocities for the HD~4203 system which confirm that the linear
trend previously observed in the residuals is due to an additional
planet. These results place dynamical constraints on the source of the
planet's eccentricities, constraints on additional planetary
companions, and informs the known distribution of multiplicity amongst
exoplanet host stars.

\end{abstract}


%%%%%%%%%%%%%%%%%%%%%%%%%%%%%%%%%%%%%%%%%%%%%%%%%%%%%%%%%%%%%%%%%%%%

\section{Introduction}
\label{intro}

Of the confirmed exoplanets, over 500 have been discovered using the
radial velocity (RV) technique. A particular attribute of the RV
systems is that they preferentially are harbored by bright host stars
due to the target selection techniques and the signal-to-noise
requirements of the surveys. The advantage of this magnitude bias is
that it enables further follow-up and characterization studies of both
the star and the harbored planets.

It has been well established that stars in the solar neighborhood have
a high rate of stellar multiplicity. This nearby multiplicity rate has
been primarily determined through spectroscopic means. \citet{abt76}
determined that binary and multiple stars systems are more common than
single stars, with a vast majority of their stellar sample showing
evidence of stellar companions. Their relatively high rate of stellar
multiplicity wer revised to a smaller value by \citet{duq91} in their
distance-complete survey of nearby solar-type stars. They found the
distribution of solar-type binaries to have a period distribution peak
near 10,000 days (30 years which corresponds to about 10 AU). A more
recent multiplicity survey by \citet{rag10} found a similar rate of
stellar companions to solar-type stars as found by \citet{duq91}.

Planets have been shown to be common from a variety of sources, such
as microlensing experiments \citep{cas12} and Kepler transiting
planets \citep{dre13}. However, relatively few of the known exoplanet
host stars have detected stellar companions and studies have indeed
found the multiplicity rate to be lower than the general stellar
population \citep{roe12}. Clearly there is an inconsistency between
the rate of stellar multiplicity of exoplanet host stars and the
general stellar population, as discussed above. The presence of
stellar companions can have significant implications for planet
formation scenarios and the subsequent properties of the systems, such
as the period-mass \citep{zuc02} and period-eccentricity \citep{egg04}
distributions. Thus it is of vital importance to clearly establish the
true rate of stellar multiplicity amongst exoplanet host stars.

\begin{deluxetable*}{lcccc}
  \tablecolumns{5}
  \tablewidth{0pc}
  \tablecaption{\label{paramtab} Star and Inner Planet Parameters}
  \tablehead{
    \colhead{Parameter} &
    \colhead{HD~4203b$^{(1)}$} &
    \colhead{HD~168443b$^{(2)}$} &
    \colhead{HD~1690b$^{(3)}$} &
    \colhead{HD~137759b$^{(4)}$}
  }
  \startdata
Distance (pc)$^{(5)}$ &
  $77.2^{+6.7}_{-5.7}$   & $37.43^{+0.99}_{-0.94}$ &
  $310^{+250}_{-96}$     & $31.027^{+0.097}_{-0.096}$ \\
$V$ mag &
  $8.70$                 & $6.92$ &
  $9.17$                 & $3.29$ \\
$[$Fe/H$]$ (dex)$^{(6)}$ &
  $0.41$                 & $0.06$ &
  $-0.32$                & $0.08$ \\
$M_\star$ ($M_\odot$) &
  $1.13^{+0.028}_{-0.1}$ & $0.995 \pm 0.019$     &
  $1.09 \pm 0.15$        & $1.4$ \\
$P$ (days) &
  $431.88 \pm 0.85$      & $58.11247 \pm 0.0003$ & 
  $533.0 \pm 1.7$        & $510.88 \pm 0.15$ \\
$T_p$ (JD - 2,440,000) &
  $11918.9 \pm 2.7$      & $15626.199 \pm 0.024$ & 
  $14449.0 \pm 5.0$      & $12013.94 \pm 0.48$ \\
$e$ &
  $0.519 \pm 0.027$      & $0.52883 \pm 0.00103$ & 
  $0.64 \pm 0.04$        & $0.7261 \pm 0.0061$ \\
$\omega$ (deg) &
  $329.1 \pm 3.0$        & $172.923 \pm 0.139$   & 
  $122.0 \pm 8.0$        & $88.7 \pm 1.4$ \\
$K$ (m/s) &
  $60.3 \pm 2.2$         & $475.133 \pm 0.9102$  &
  $190.0 \pm 29.0$       & $299.9 \pm 4.3$ \\
$M_p \sin i$ ($M_J$) &
  $2.08 \pm 0.116$       & $7.66 \pm 0.098$    &
  $6.1 \pm 0.9$          & $10.3$ \\
$a$ (AU) &
  $1.165 \pm 0.022$      & $0.2931 \pm 0.00181$  & 
  $1.30 \pm 0.02$        & $1.34$ \\
Slope (m/s/yr) &
  $-4.38 \pm 0.71$       & $-3.17 \pm 0.09$ &
  $-7.2 \pm 0.4$         & $-13.8 \pm 1.1$
  \enddata
  \tablenotetext{(1)}{\citet{but06}}
  \tablenotetext{(2)}{\citet{pil11}}
  \tablenotetext{(3)}{\citet{mou11}}
  \tablenotetext{(4)}{\citet{zec08}}
  \tablenotetext{(5)}{\citet{van07}}
  \tablenotetext{(6)}{See Section \ref{abundances}}
\end{deluxetable*}

There have been several previous studies that have conducted searches
for stellar companions. \citet{ber13} used lucky imaging to conduct
such a search around relatively faint transiting planet host
stars. The resolution of their observations allowed the detection of
companion 4 magnitudes fainter at $0.5\arcsec$
separation. \citet{egg07} conducted a survey of southern exoplanet
host stars using VLT/NACO and were able to detect a handful of
previously unknown companions to some of the exoplanet host
stars. Attempts are also underway to detect stellar companions to
exoplanet host stars that exhibit a linear trend in the RV data
\citep{cre12,cre13}.

Here we present the results of high-resolution imaging of four
exoplanet host stars with known RV linear trends in the residuals in
order to detect or constrain the presence of stellar companions.  We
also present new Keck/HIRES RV data for HD~4203 which confirms a
second planet in that system. Imaging observations were carried out
using the Differential Speckle Survey Instrument (DSSI) on
Gemini-North, a system which has already been successfully used to
observe much fainter stars in the Kepler and CoRoT fields
\citep{hor12}. Each of the four target systems in this study harbor a
giant planet in an eccentric orbit whose eccentricity may in part be
explained by the perturbing influence of the third body. Two of the
stars are giants and two are dwarfs. In Section 2 we describe in
detail the science motivation behind the selection of the four targets
and their relevant properties. In Section 3 we present new RV data
which confirms that the linear trend for HD~4203 is due to an
additional planet in that system. Section 4 outlines the DSSI
observations and data reduction processes. In Section 5 we present the
results of the imaging data and subsequent constraints on companions
in terms of $\Delta$~magnitude and angular separation. We map these
results in Section 6 to constraints on companions in terms of stellar
mass and physical separation. Section 7 briefly describes stellar
abundances for the targets in the context of stellar companions and
exoplanet orbital eccentricity. Finally, we provide concluding remarks
in Section 8, commenting on the implications for limits on stellar and
planetary companions in these system.

%%%%%%%%%%%%%%%%%%%%%%%%%%%%%%%%%%%%%%%%%%%%%%%%%%%%%%%%%%%%%%%%%%%%

\section{Target Selection}

Our observations were scheduled to occur during the 2013B semester on
Gemini North, during which the DSSI instrument (described in Section
\ref{observations}) was deployed as a visiting instrument. We thus
selected targets within a Right Ascension range of $15h < RA < 03h$
and a Declination range of $-10\degr < Dec < +70\degr$. The targets in
our list have all been the subject of radial velocity studies from
which their known planetary companions were detected. We elected to
focus our observations on bright stars ($V < 12$) with at least one
planet in a highly eccentric ($e > 0.5$) orbit. Furthermore, we
required that there be no known binary companion to the host star but
a radial velocity trend present in the residuals of the Keplerian
planetary fit to the data. The above criteria were applied to the
known stellar and planetary parameters using the data stored in the
Exoplanet Data Explorer\footnote{\tt http://exoplanets.org/}
\citep{wri11}. This resulted in our final list of targets which
include HD~4203, HD~168443, HD~1690, and HD~137759 (iota Draconis).

\begin{figure}
  \includegraphics[angle=270,width=8.2cm]{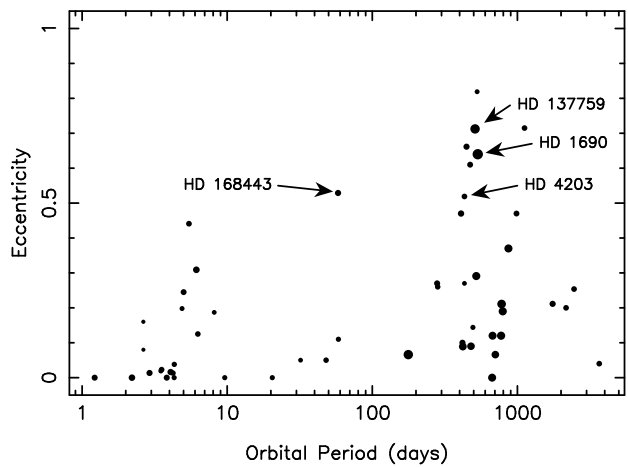}
  \caption{A plot of eccentricity versus orbital period (inner planet)
    for all systems which are known to have a radial velocity trend in
    the Keplerian orbital solution and are not known to have a stellar
    companion to the host star. The size of each plotted point is
    logarithmically proportional to the radius of the host star to
    distinguish dwarfs from giants. The locations of the four systems
    studied in this paper are labelled.}
  \label{targetfig}
\end{figure}

\begin{figure*}
  \begin{center}
    \begin{tabular}{cc}
      \includegraphics[width=8.2cm]{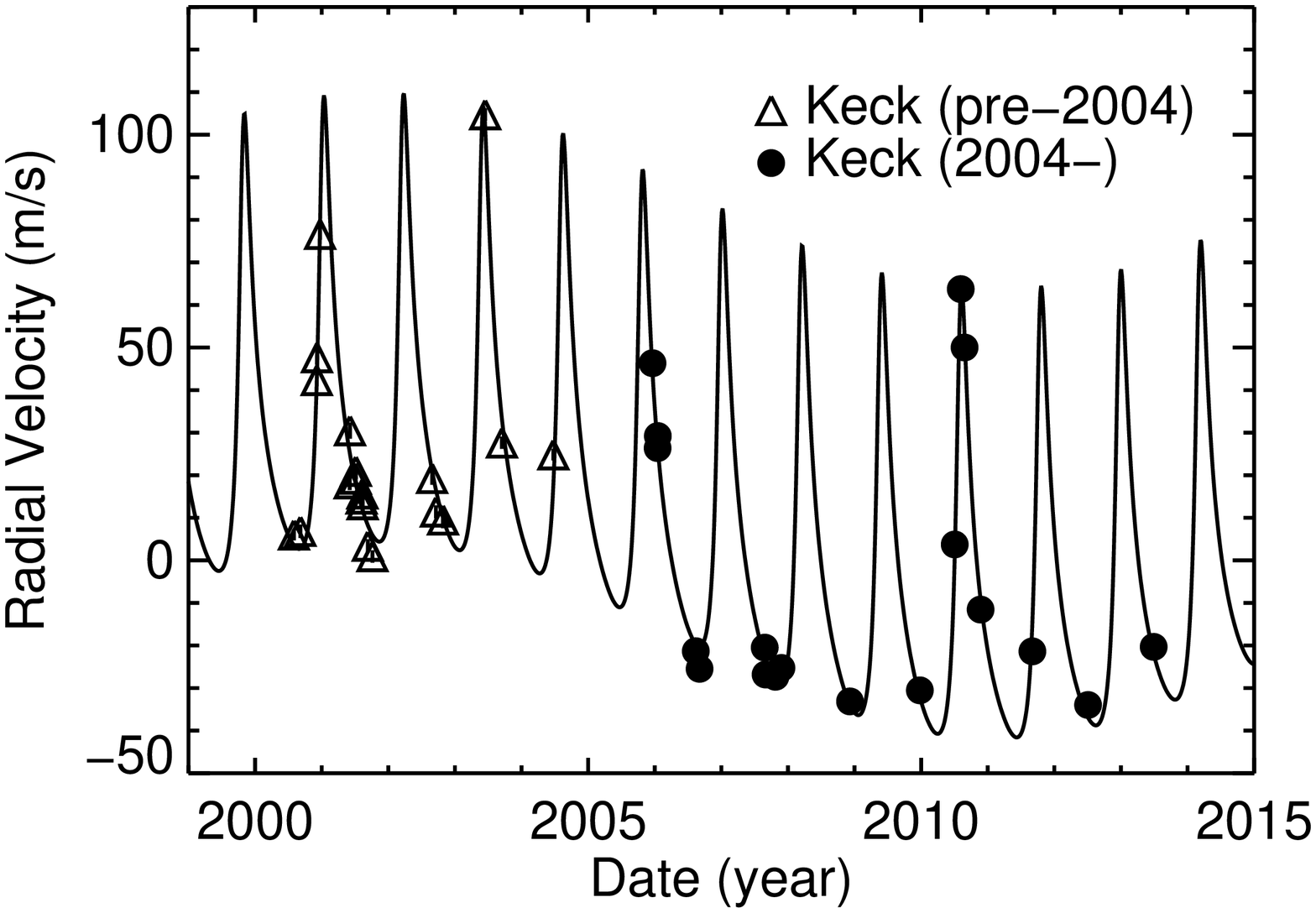} &
      \includegraphics[width=8.2cm]{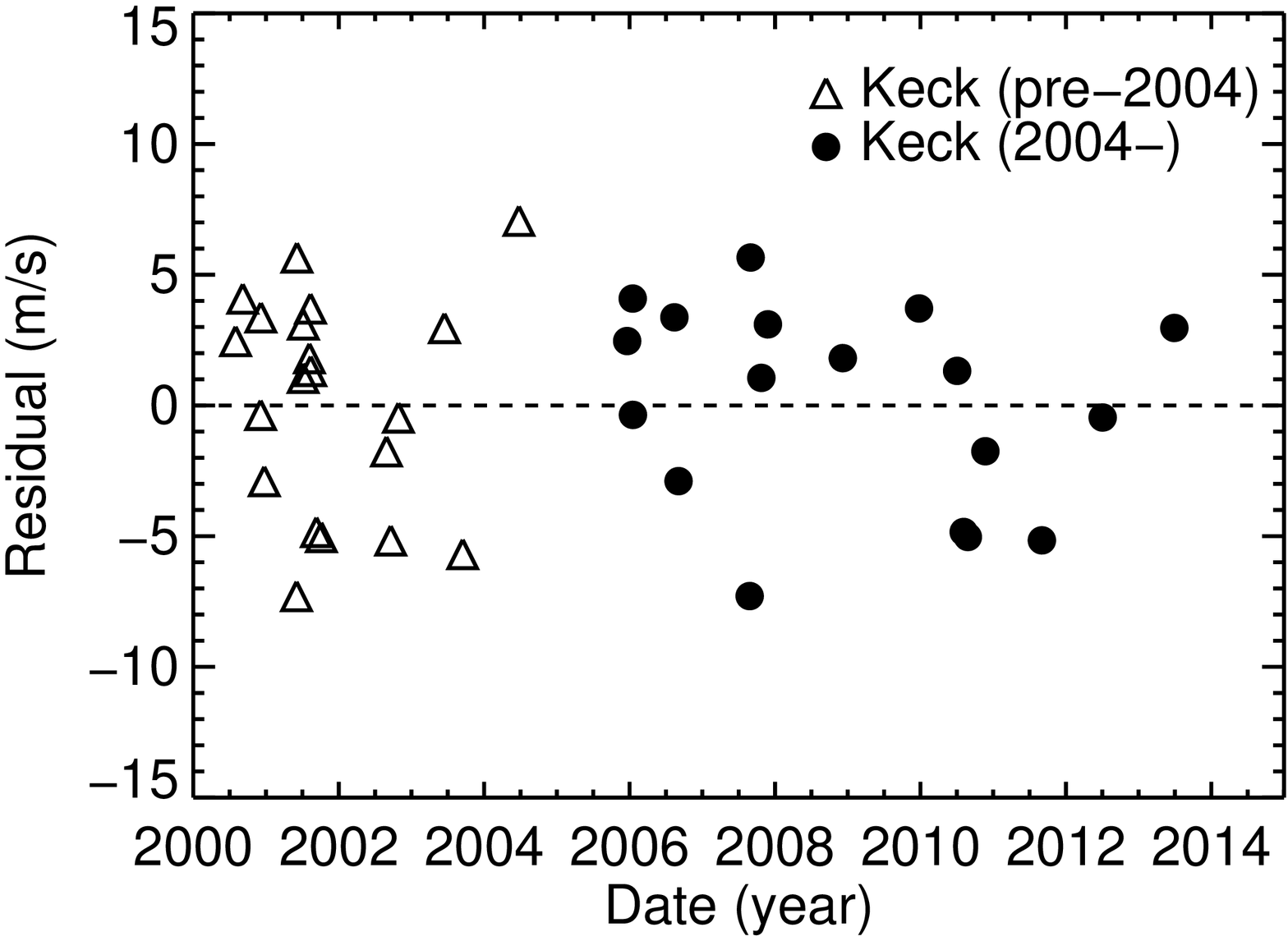} \\
      \includegraphics[width=8.2cm]{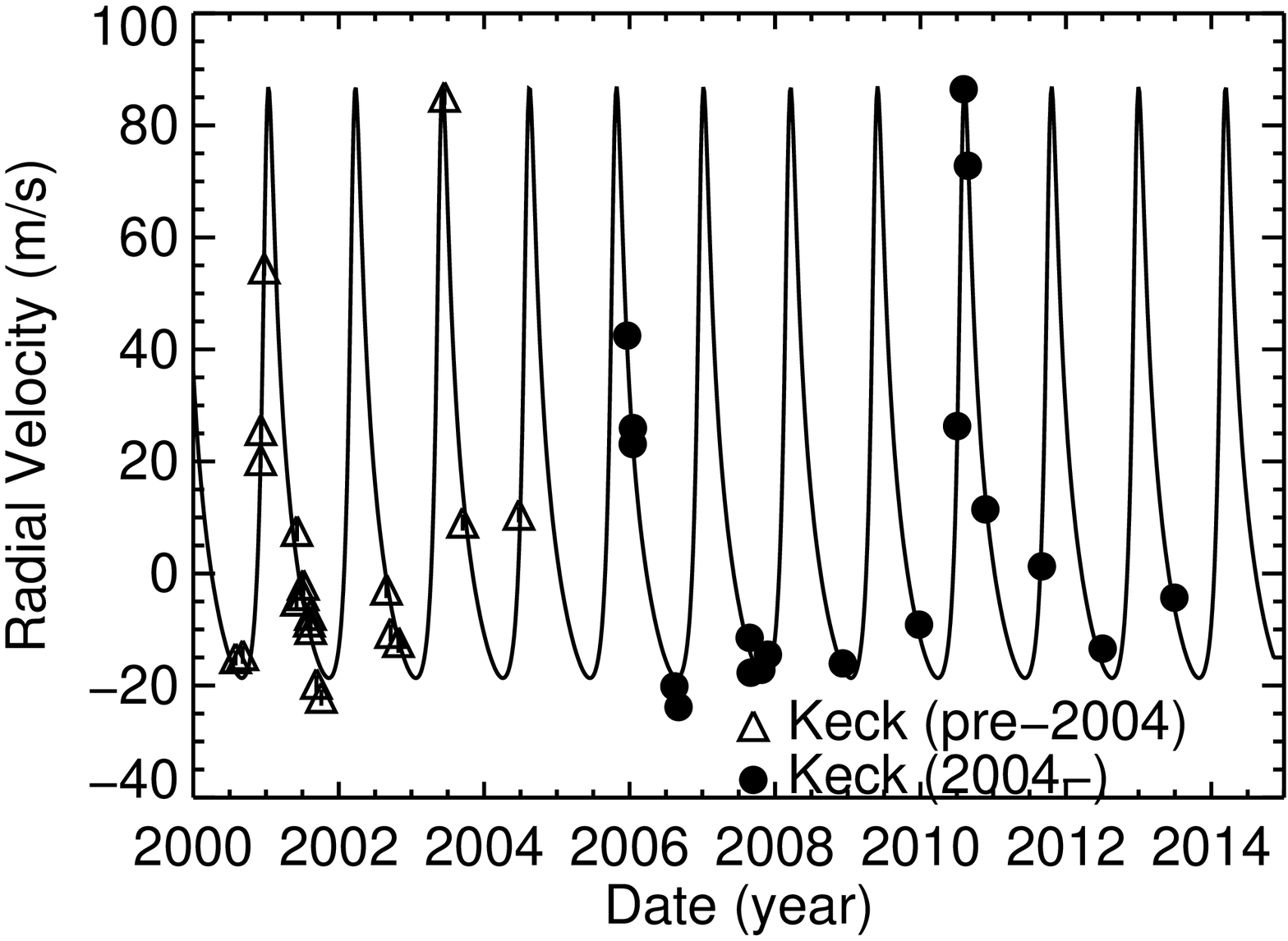} &
      \includegraphics[width=8.2cm]{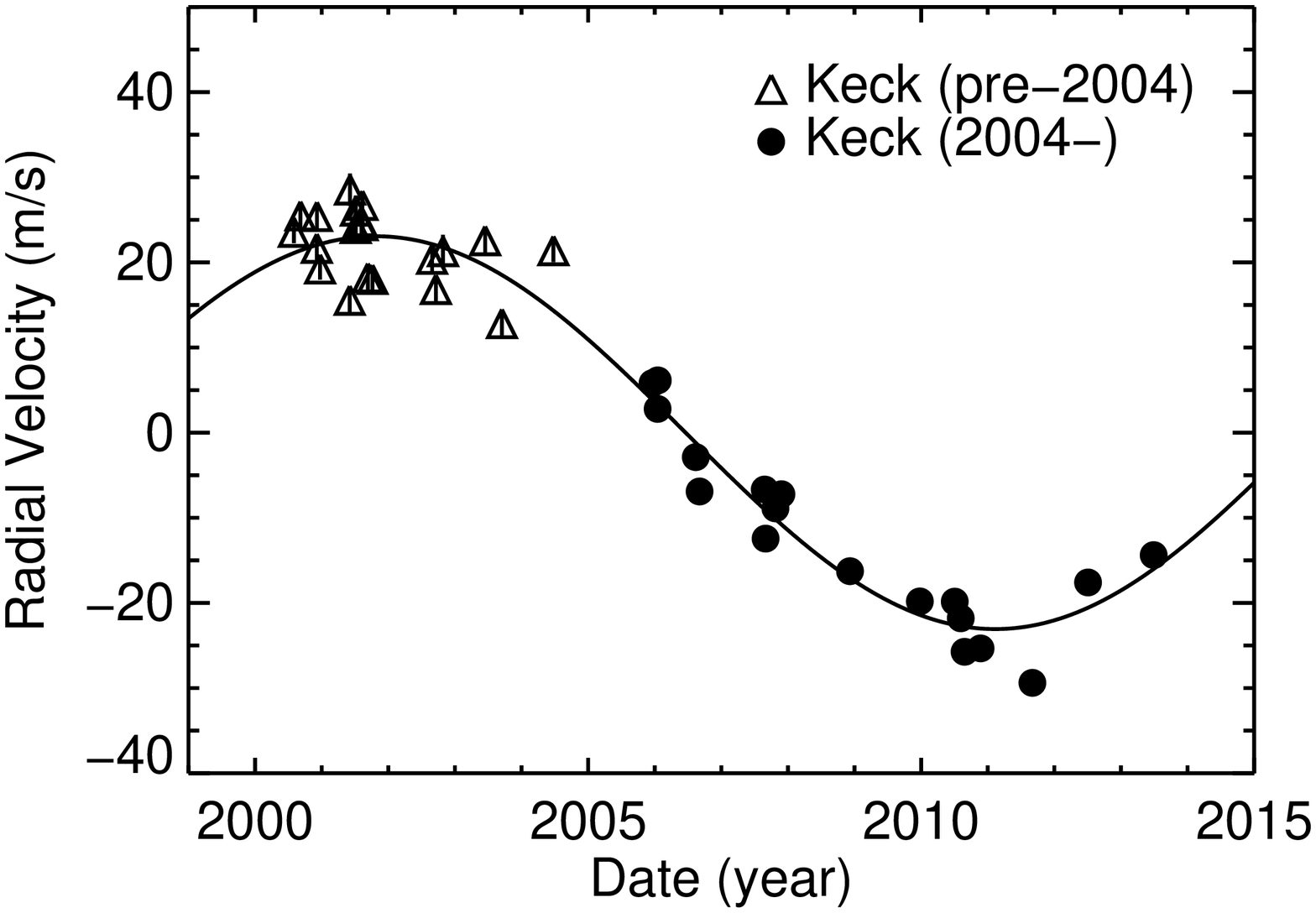}
    \end{tabular}
  \end{center}
  \caption{Radial velocity and Keplerian fits for the HD 4203
    system. In all plots, triangles represent pre-2004 Keck RVs and
    filled circles represent post-2004 Keck RVs. Solid lines represent
    the best-fit Keplerian orbits. Top-left: Keck RVs overplotted by
    best-fit 2-planet Keplerian model. Top-right: Residuals of the RVs
    with the best-fit 2-planet Keplerian model
    subtracted. Bottom-left/bottom-right: The RV curves for HD 4203 b
    and c, respectively. We implemented a best-fit offset of 2.38 m/s
    in plotting these figures.}
  \label{4203plot}
\end{figure*}

The Keplerian orbital solutions for the four planets/systems along
with relevant stellar parameters and references are shown in Table
\ref{paramtab}. Note that the HD~168443 system has two known planets
and thus the Keplerian orbital solution applies to the b planet. The
HD~4203 system also has two planets which are described further in
Section \ref{hd4203update}. The effective search radius for DSSI is
$0.05\arcsec$ to $1.4\arcsec$. This means, for example, stars located
at 30 pc have a search radius of 1.5--42~AU. As noted above, we
selected stars with eccentric planets since a primary motivation is to
explore the possibility of scenarios which include a perturbing body
which has exchanged angular momentum from the eccentric planet. We
also chose to select two dwarf stars (HD~4203 and HD~168443) and two
giant stars (HD~1690 and HD~137759) to investigate these scenarios for
a range of luminosity classes. Many of these criteria are summarized
in Figure \ref{targetfig}, which shows a plot of orbital eccentricity
versus orbital period (of the inner planet) for those systems which
have a known radial velocity trend but no known stellar companion to
the host star. The size of each data point is logarithmically
proportional to the radius of the host star. The four targets selected
for observations are appropriately labelled and shows that many of the
eccentric planets in this population fall in the 400--1000~day period
range. The HD~168443 system is an outlier in this plot and thus
represents a significantly different system architecture to the other
three.

%%%%%%%%%%%%%%%%%%%%%%%%%%%%%%%%%%%%%%%%%%%%%%%%%%%%%%%%%%%%%%%%%%%%

\section{An Update to the HD 4203 Planetary System}
\label{hd4203update}

We present an updated fit to HD 4203 b with additional radial
velocities (RVs) from HIRES at the Keck Observatory
\citep{vog94}. Since the work of \citet{vog02} and \citet{but06}, we
have monitored HD 4203 up to mid-2013. Table \ref{4203rv} lists the 38
HIRES RV observations. We found that the previously identified trend
was due to an outer planet. HD 4203c has a period of $\sim 18.2$ years
and a minimum mass of $\sim 2.17 M_{\rm Jup}$. Table \ref{4203} lists
the orbital parameters of HD 4203b and HD 4203c. We fit Keplerian
orbits to the RV data with the RVLIN package \citep{wri09}. We applied
a stellar jitter of 4m/s \citet{but06}. We derived uncertainties for
the parameters using BOOTTRAN, the bootstrapping package described by
\citet{wan12}. The combined system fit, RV residuals, and fits for
each planet are shown in Figure \ref{4203plot}.

\begin{figure}
  \includegraphics[width=8.2cm]{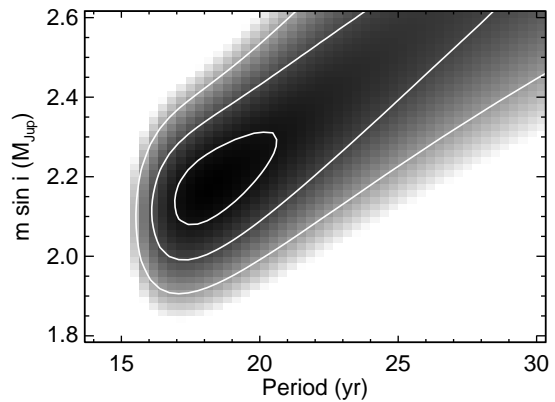}
  \caption{Best-fit $50 \times 50$ $\chi^2$ map for fixed values of
    \pc and \mc, with offset penalty applied. This confirms that the
    period and mass are constrained to $1\sigma$. We have illustrated
    the contours of the $1\sigma$, $2\sigma$, and $3\sigma$ (defined
    by $\chi^2=\chi^2_{\rm min} +\{2.30,6.17,11.8\}$) confidence
    levels. }
  \label{chi2}
\end{figure}

Because the orbit of HD 4203c has not completed, BOOTTRAN by itself
could not constrain well the period, $P_{c}$, and minimum mass, $M_{c}
\sin{i}_{c}$. We also treated the RVs taken before and after the HIRES
upgrade in 2004 separately, which results in an offset between the two
streams of velocities. In order to constrain the period and mass, we
constructed a \kai map for the best-fit Keplerian orbits where we
fixed values of \pc and \mc at each point in the map. We let all other
parameters float, giving us a total of 10 free parameters: 5 for
planet \textit{b}, 3 for planet \textit{c}, $\gamma$ for the system
and offset between the two RV streams. When the offset was a floating
parameter in our fitting, we found parameters were not well
constrained. Therefore we enforced a penalty for the values of offset
from our fits to obtain the \kai map shown in Figure \ref{chi2}.

\begin{figure}
  \includegraphics[width=8.2cm]{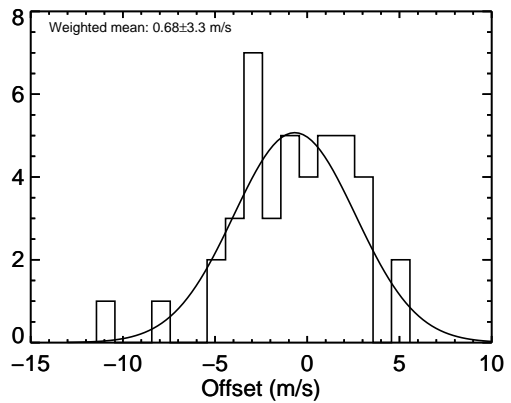}
  \caption{The histogram distribution of offsets for 42 stars from the
    Keck catalog. We selected the stars based on similarity to HD
    4203, where $\Delta T_{\rm eff}< 225$ K and $\Delta \log{g} <
    0.3$. We excluded stars which exhibited trends or are known hosts
    of long-period ($>1$ year) planets, and stars that have less than
    10 observations for both the pre- and post-2004 RV streams. We
    calculated the offset for each star's set of RVs and found a
    weighted mean of $-0.68$ m/s and a variance of 10.91. We overplot
    a Gaussian with the same mean and variance.}
  \label{offsethist}
\end{figure}

We first had to determine the offset penalty term. To do so, we
identified a pool of Keck stars with similar temperature and gravity
as HD 4203 using $\Delta T_{\rm eff}= 225$ K and $\Delta \log{g} =
0.3$. Within that pool, we filtered out the stars with less than 10
observations for both pre- and post-2004 RVs. We then examined the RV
curve of each star to exclude stars that have significant trend or
known long-period planets. For stars with planets that are in short
orbits ($< 1$ year), we subtract the orbital solution. We calculated
the offset as the difference in mean of the pre- and post-2004
streams. This selection yielded a list of 42 stars for which we
calculated the individual offsets and the weighted mean and variance
of all the offsets. We plot the distribution of offsets from our list
of 42 stars (see Figure \ref{offsethist}).

\begin{deluxetable}{ccc}
\tablecaption{Radial velocities for HD 4203 from Keck \label{4203rv}}
\tablewidth{0pt}
\tablehead{
\colhead{Time} & \colhead{Velocity} & \colhead{Uncertainty}\\
\colhead{(JD - 2,440,000)} & \colhead{(m/s)} & \colhead{(m/s)}
}
\startdata
       11757.122 &        6.0021463 &       1.31503 \\
       11792.972 &        6.8441796 &       1.58321 \\
       11882.835 &        42.293574 &       1.39100 \\
       11883.848 &        47.667231 &       1.63549 \\
       11900.838 &        76.619395 &       1.28310 \\
       12063.126 &        18.041178 &       1.52620 \\
       12065.129 &        30.488276 &       1.90241 \\
       12096.114 &        19.101145 &       1.47675 \\
       12097.068 &        20.965464 &       1.80766 \\
       12128.116 &        14.444193 &       1.65465 \\
       12133.056 &        15.628565 &       1.49746 \\
       12133.926 &        13.116808 &       1.57353 \\
       12162.918 &        3.3657251 &       1.56790 \\
       12187.962 &       0.90439083 &       1.44404 \\
       12515.017 &        19.283503 &       1.52212 \\
       12535.989 &        11.405924 &       1.52340 \\
       12574.784 &        9.5314431 &       1.95447 \\
       12806.126 &        104.74114 &       1.46572 \\
       12897.903 &        27.623347 &       1.43196 \\
       13180.122 &        24.839053 &       1.36543 \\
       13723.747 &        48.697809 &      0.935456 \\
       13751.749 &        28.752323 &      0.899553 \\
       13752.814 &        31.586298 &      0.921903 \\
       13961.019 &       -18.960217 &       1.02713 \\
       13981.909 &       -23.106548 &      0.888949 \\
       14339.103 &       -18.116657 &       1.00089 \\
       14343.931 &       -24.443002 &      0.956635 \\
       14397.802 &       -24.972989 &      0.961509 \\
       14430.763 &       -22.894478 &      0.985528 \\
       14806.840 &       -30.759572 &       1.13911 \\
       15189.785 &       -28.156682 &      0.965326 \\
       15381.079 &        6.1080502 &      0.933918 \\
       15414.027 &        66.112621 &       1.01668 \\
       15435.044 &        52.387934 &       1.03033 \\
       15522.893 &       -9.1774166 &       1.05714 \\
       15806.918 &       -19.001478 &      0.988947 \\
       16112.131 &       -31.595639 &      0.991987 \\
       16472.128 &       -17.945240 &      0.939638 \\
\enddata
\end{deluxetable}

\begin{deluxetable}{llccc}
  \tablecaption{Orbital parameters of the HD 4203 planetary
    system \label{4203}}
  \tablewidth{0pt}
  \tablehead{
    \multicolumn{2}{c}{Parameter} & \colhead{} &\colhead{HD 4203 b} & \colhead{HD 4203 c}
  }
  \startdata
  $P$ & days              &  & $437.05 \pm 0.27$   & $6700 \pm 4500$ \\
  $T_p$ & JD - 2,440,000  &  & $11911.52 \pm 2.38$ & $16000 \pm 9600$\\
  $e$ &                   &  & $0.52 \pm 0.02$     & $0.24\pm 0.13$\\
  $\omega$ & (deg)        &  & $328.03 \pm 2.9$    & $224 \pm 48.8$ \\
  $K$ & m/s               &  & $52.82 \pm 1.5$     & $22.20 \pm 3.707$\\
  $M_p \sin i$ & ($M_J$)  &  & $1.82 \pm 0.05$     & $2.17 \pm 0.52$\\
  $a$ & (AU)              &  & $1.1735 \pm 0.0222$ & $6.95^{+1.93}_{-0.56}$\\
  \hline
  $\rm RMS$ & m/s         &  & \multicolumn{2}{c}{3.93}\\
  $\chi^2_{\nu}$ &        &  & \multicolumn{2}{c}{0.87}
  \enddata
\end{deluxetable}

\begin{deluxetable*}{lcccccc}
  \tablecolumns{7}
  \tablewidth{0pc}
  \tablecaption{\label{resultstab} Results of DSSI observations}
  \tablehead{
    \colhead{Star} &
    \multicolumn{2}{c}{Exclusion radius (AU)} &
    \multicolumn{2}{c}{$5\sigma$ $\Delta m$ Limit (692~nm)} &
    \multicolumn{2}{c}{$5\sigma$ $\Delta m$ Limit (880~nm)} \\
    \colhead{} &
    \colhead{Inner} &
    \colhead{Outer} &
    \colhead{$0.1\arcsec$} &
    \colhead{$0.2\arcsec$} &
    \colhead{$0.1\arcsec$} &
    \colhead{$0.2\arcsec$}
  }
  \startdata
HD~4203   &  3.86 & 108.1 & 4.87 & 5.30 & 4.53 & 5.20 \\
HD~168443 &  1.87 &  52.4 & 2.93 & 4.55 & 4.19 & 5.17 \\
HD~1690   & 15.50 & 434.0 & 4.16 & 4.64 & 3.51 & 4.59 \\
HD~137759 &  1.55 &  43.4 & 4.33 & 4.79 & 3.78 & 4.64
  \enddata
\end{deluxetable*}

We created the offset-penalized \kai map from an array of 101 offset
values between -10 m/s and 10 m/s. Each offset had a corresponding 50
by 50 \kai map. Next, we applied each offset to the series of RVs
taken after 2004. The penalty term we add to each map is given by
$(o-\mu_{\rm w})^2/\sigma_{o}^2$, where $o$ is the corresponding
offset, and $\mu_{\rm w}$ and $\sigma_{o}^2$ are the weighted mean and
variance of the offsets from the selected set of stars. For each
pixel, we selected the minimum \kai value from the stack of 101
offsets. The minima for all the stacks made up the offset-penalized
\kai map. We also marked contours of the $1\sigma$, $2\sigma$, and
$3\sigma$ confidence levels (calculated as $\chi^2=\chi^2_{\rm min}
+\{2.30,6.17,11.8\}$) for the 2D \kai distribution \citep{pre02}.

In addition to each of the 101 maps, we ran BOOTTRAN. Every orbital
parameter had a structure 1000 by 101 in dimension from the BOOTTRAN
runs and the range of offsets. We implemented a two-level mask that
first randomly excluded points less than the weight set by the offset
distribution in Figure \ref{offsethist}, normalized to 1. Secondly, we
masked out fits that returned a period of 100 years, which indicates
that the fit failed. Using the mask, we created distributions of values
for each parameter and determined the width, which was used for the
uncertainty. The minimum mass required separate calculation, and we
generated an array of \mc based on the distributions of period,
velocity semiamplitude, eccentricity, and mass of the star. In Table
\ref{4203}, we report the uncertainties obtained for HD 4203c.

%%%%%%%%%%%%%%%%%%%%%%%%%%%%%%%%%%%%%%%%%%%%%%%%%%%%%%%%%%%%%%%%%%%%

\section{DSSI Observations and Data Reduction}
\label{observations}

The DSSI camera is a dual-channel speckle imaging system where each
channel records speckle patterns in a different filter
\citep{hor09}. A dichroic beamsplitter splits the white light that
enters the instrument into two wavelength regimes, and then two
filters further tailor the bandpasses to the desired center
wavelengths and widths. In the case of the observations described
here, the two filters used were a red filter centered on 692 nm with a
40 nm full width at half maximum (FWHM), and a near-infrared filter
centered on 880 nm with a 50 nm FHWM. This set-up is the same as
described in \citet{hor12}; further information regarding the
magnification and the technique for measuring the pixel scale is given
there.

Observations for our targets occurred on 24--25, 27--28 July 2013.
Each target was observed by taking a sequence of 1000 60-ms frames
that were recorded simultaneously in each filter. The pixel format was
$256\times256$. These data were then processed using the image
reconstruction algorithms previously developed for the DSSI camera and
described most recently in \citet{how11}. Briefly, the method is
Fourier-based; we begin by computing the autocorrelation of the data
frames, averaging these, and then Fourier transforming the result. The
square-root of this can be divided by the same result for a point
source to arrive at the the modulus of the object's Fourier
transform. We also compute the so-called ``near-axis'' subplanes of
the bispectrum (see \citet{loh83}), which contain information about
the derivative of the object's phase in the Fourier plane. We obtain
the phase using the relaxation technique of \citet{men90}. The modulus
and phase functions are then combined and low-pass filtered with a
Gaussian filter in the Fourier plane, and the final result is
inverse-transformed to give the final reconstructed image.

The goal of the DSSI observations of the objects discussed here is the
detection of companion stars with sub-arcsecond separations relative
to the target. The objects observed were found to be single to the
limit of our detection capabilities with DSSI; to estimate the
limiting magnitude as a function of separation from the target, we
used the basic methodology described in \citet{hor11}, though the
image reconstruction routines have been improved somewhat since that
work was completed. The result is a 5$\sigma$ detection curve, showing
the limiting magnitude as a function of separation, generally with
increasing magnitude difference limits for larger separations.

%%%%%%%%%%%%%%%%%%%%%%%%%%%%%%%%%%%%%%%%%%%%%%%%%%%%%%%%%%%%%%%%%%%%

\section{Imaging Results}
\label{imagingresults}

\begin{figure*}
  \begin{center}
    \begin{tabular}{cc}
      \includegraphics[width=5.2cm]{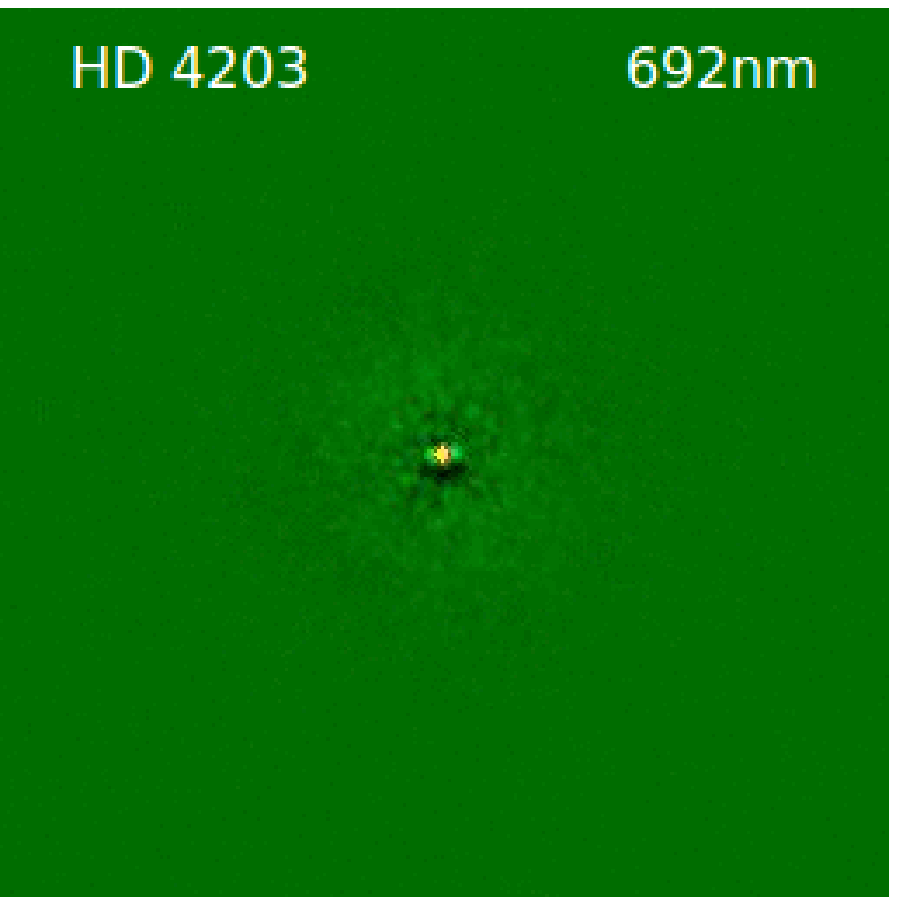} &
      \includegraphics[width=5.2cm]{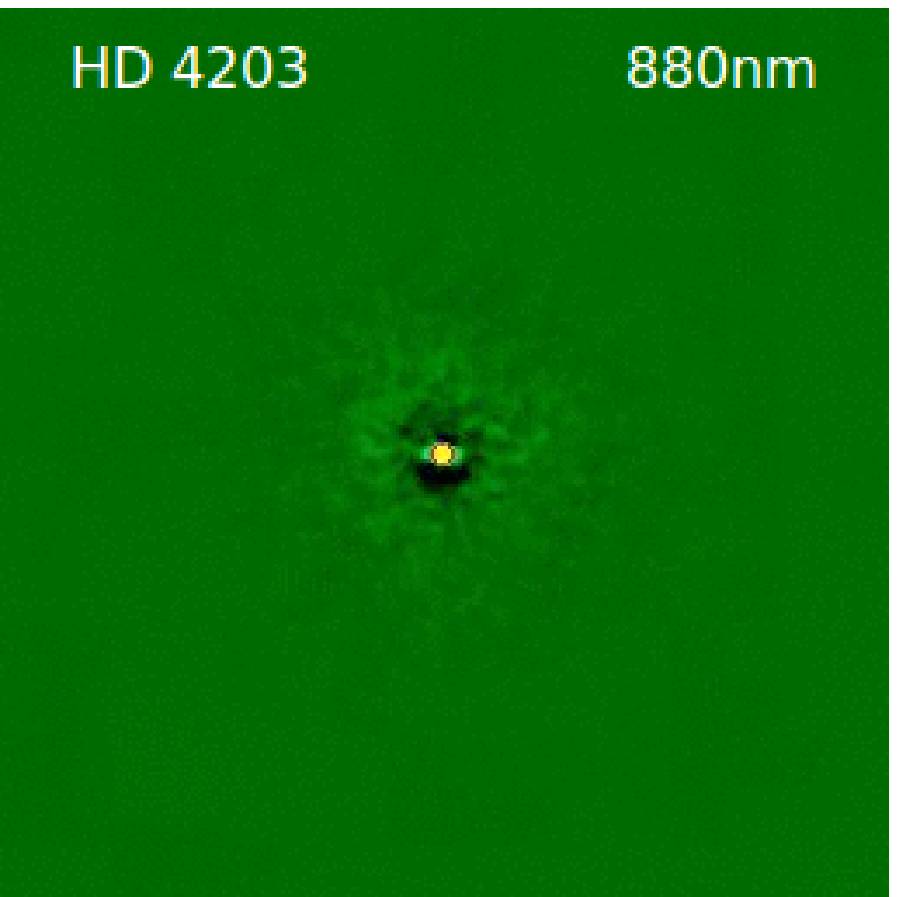} \\
      \includegraphics[width=5.2cm]{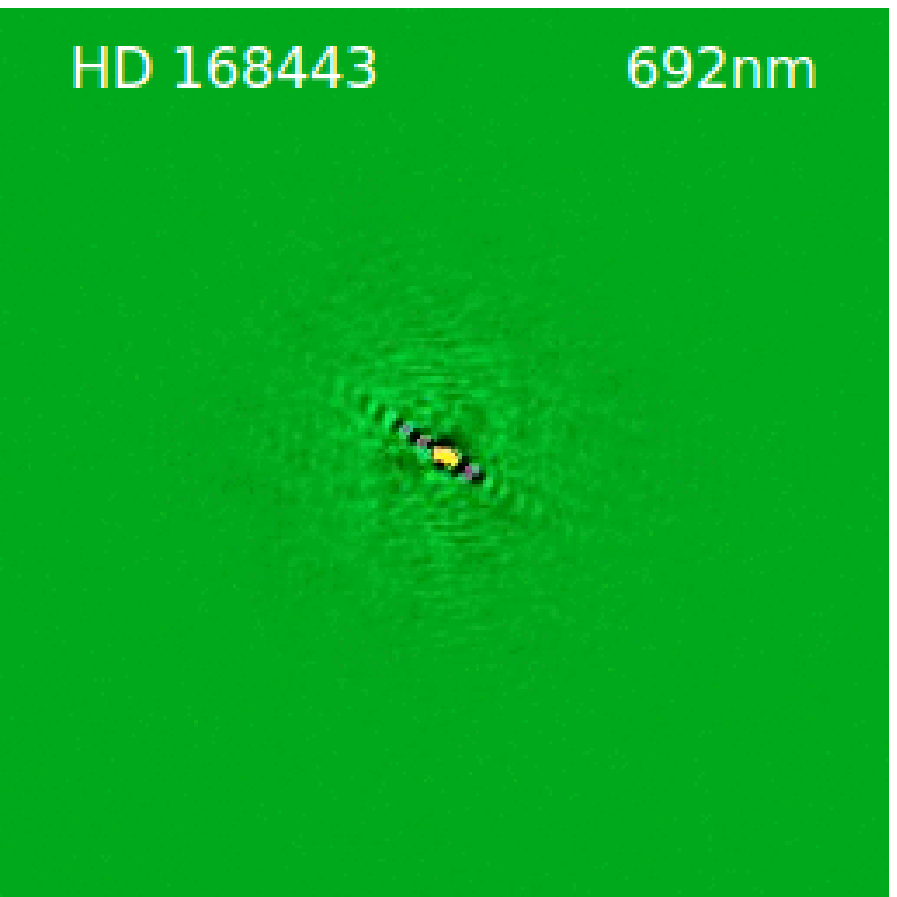} &
      \includegraphics[width=5.2cm]{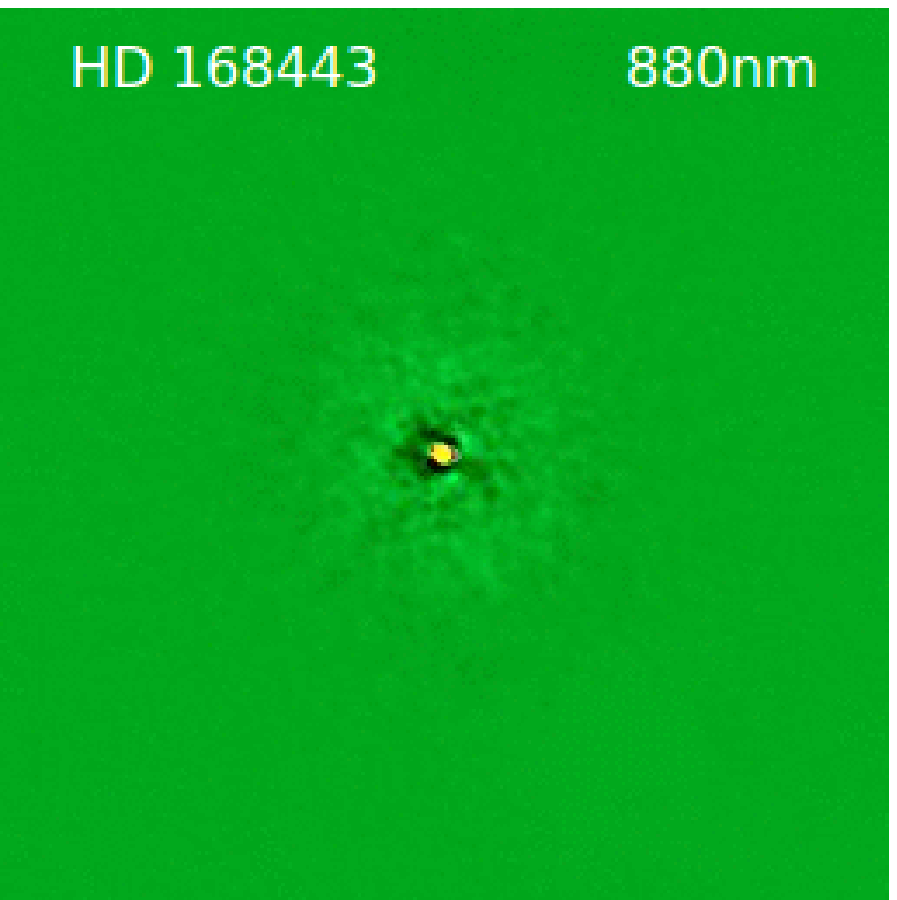} \\
      \includegraphics[width=5.2cm]{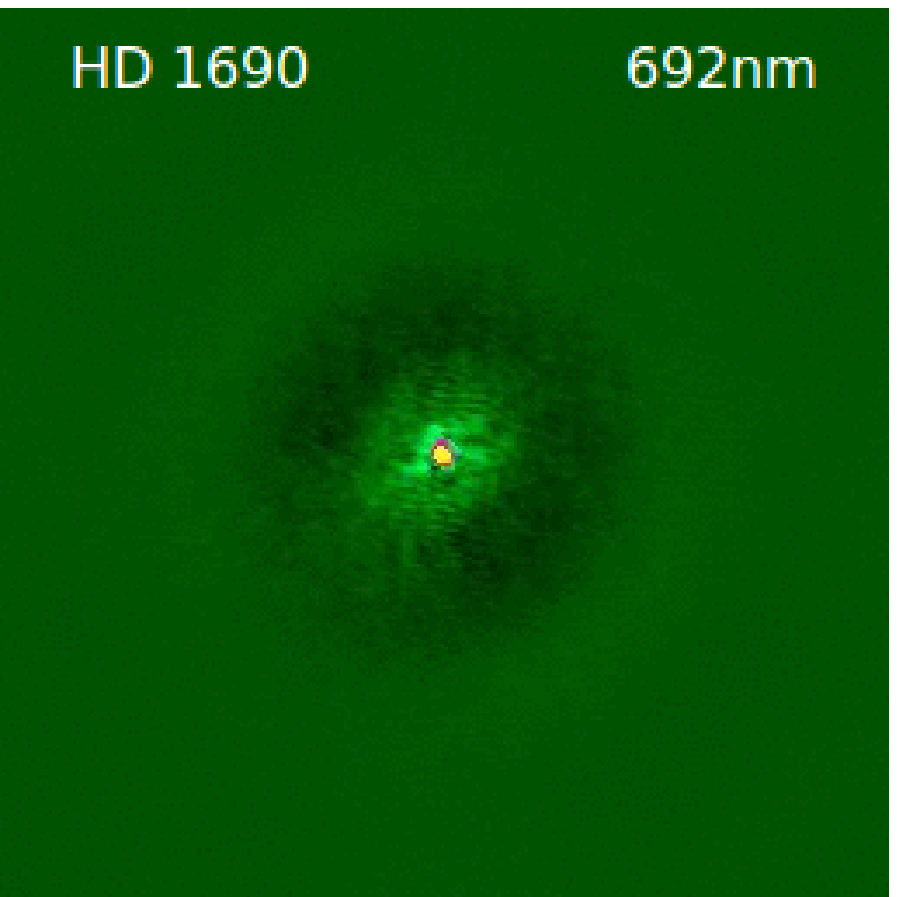} &
      \includegraphics[width=5.2cm]{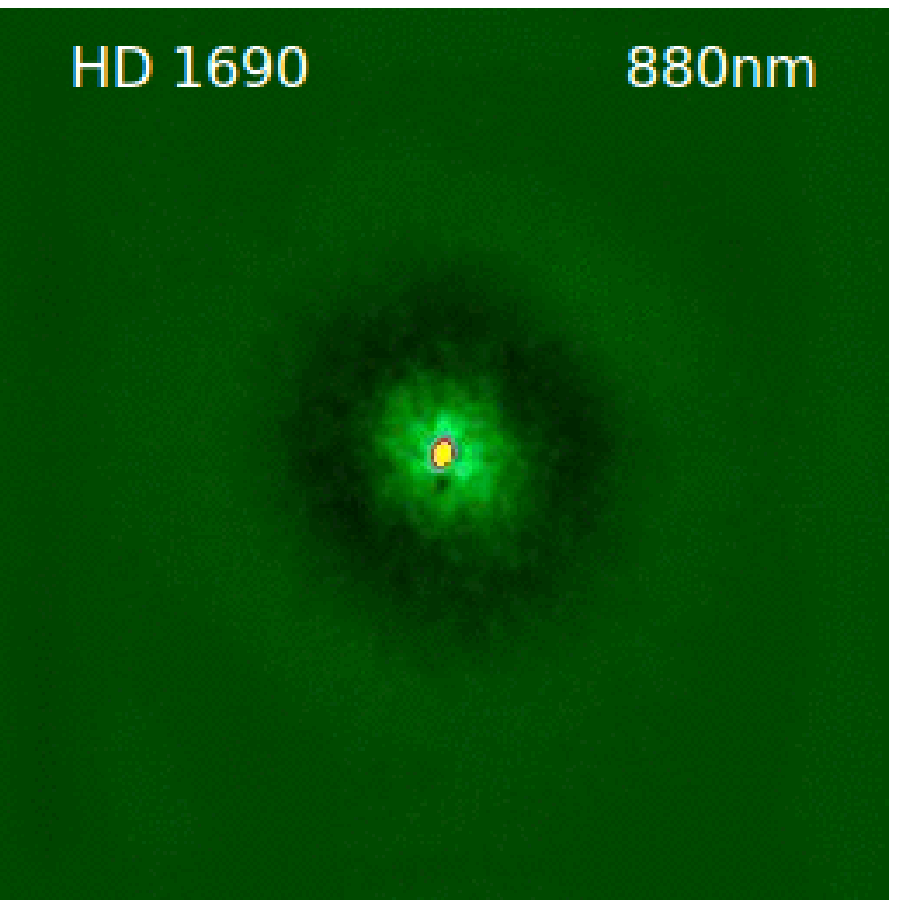} \\
      \includegraphics[width=5.2cm]{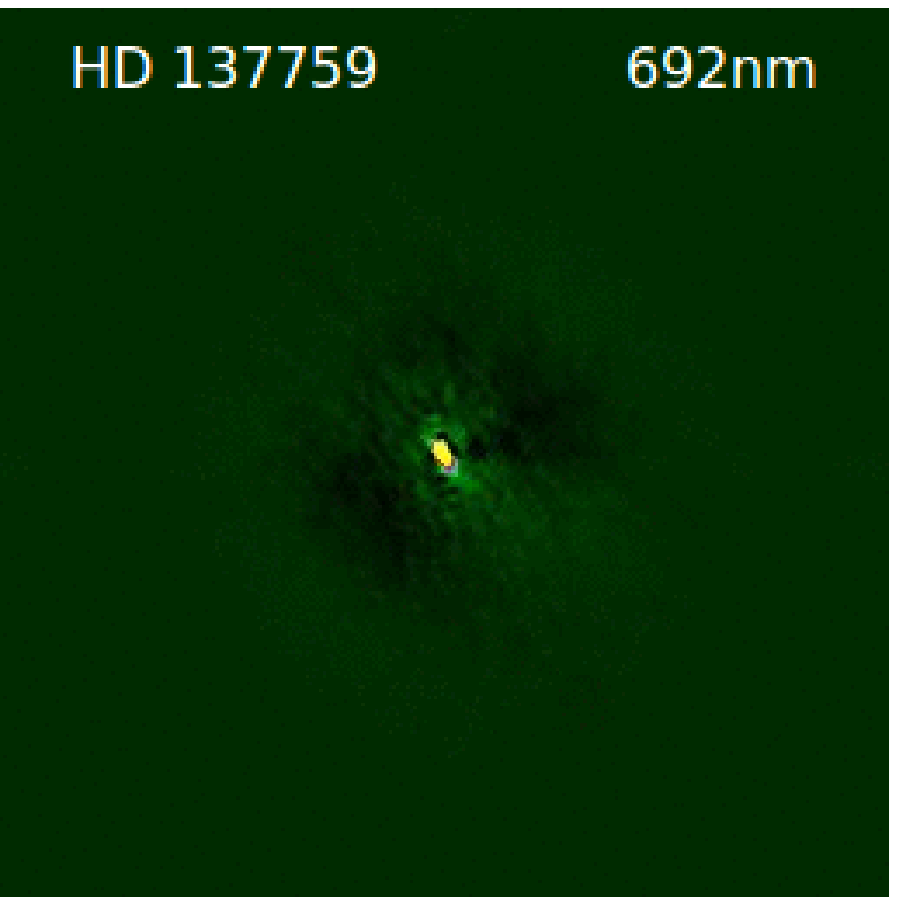} &
      \includegraphics[width=5.2cm]{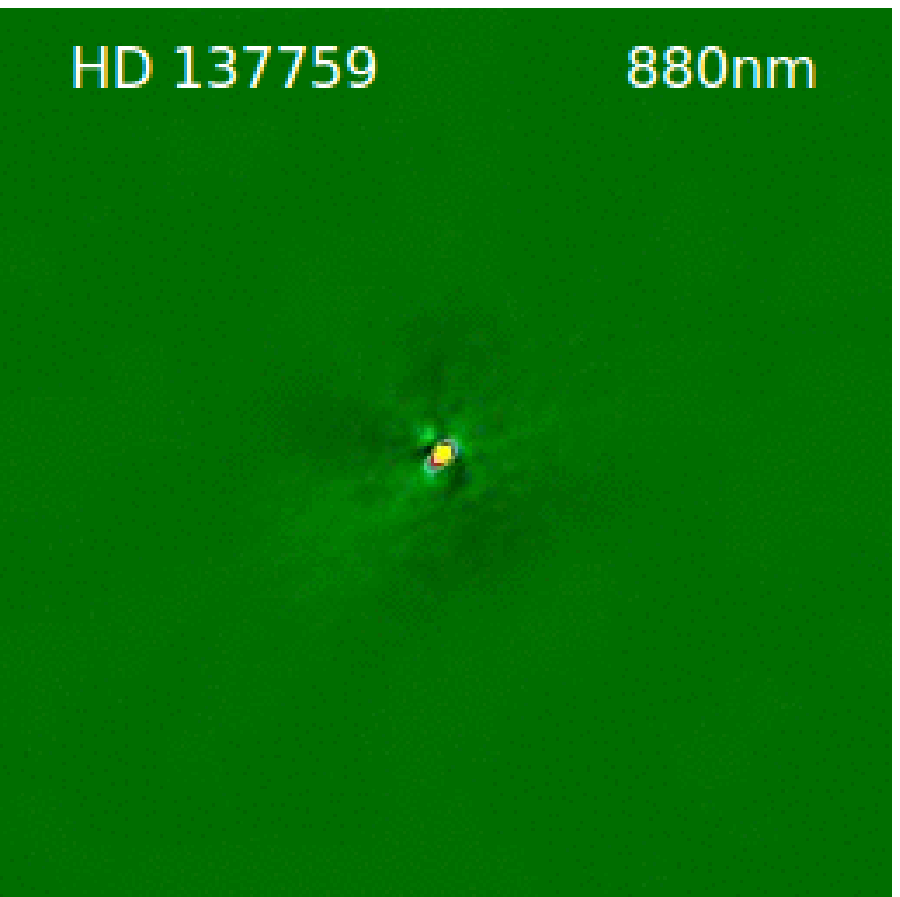} \\
    \end{tabular}
  \end{center}
  \caption{Gemini DSSI speckle images of HD~4203, HD~168443, HD~1690,
    and HD~137759 (iota Draconis). The left and right columns show the
    692~nm and 880~nm data respectively. The field-of-view is
    2.8$\times$2.8$\arcsec$. For the 692~nm images, North is up and
    East is to the right. For the 880~nm images, North is up and East
    is to the left.}
  \label{imagesfig}
\end{figure*}

\begin{figure*}
  \begin{center}
    \begin{tabular}{cc}
      \includegraphics[width=7.2cm]{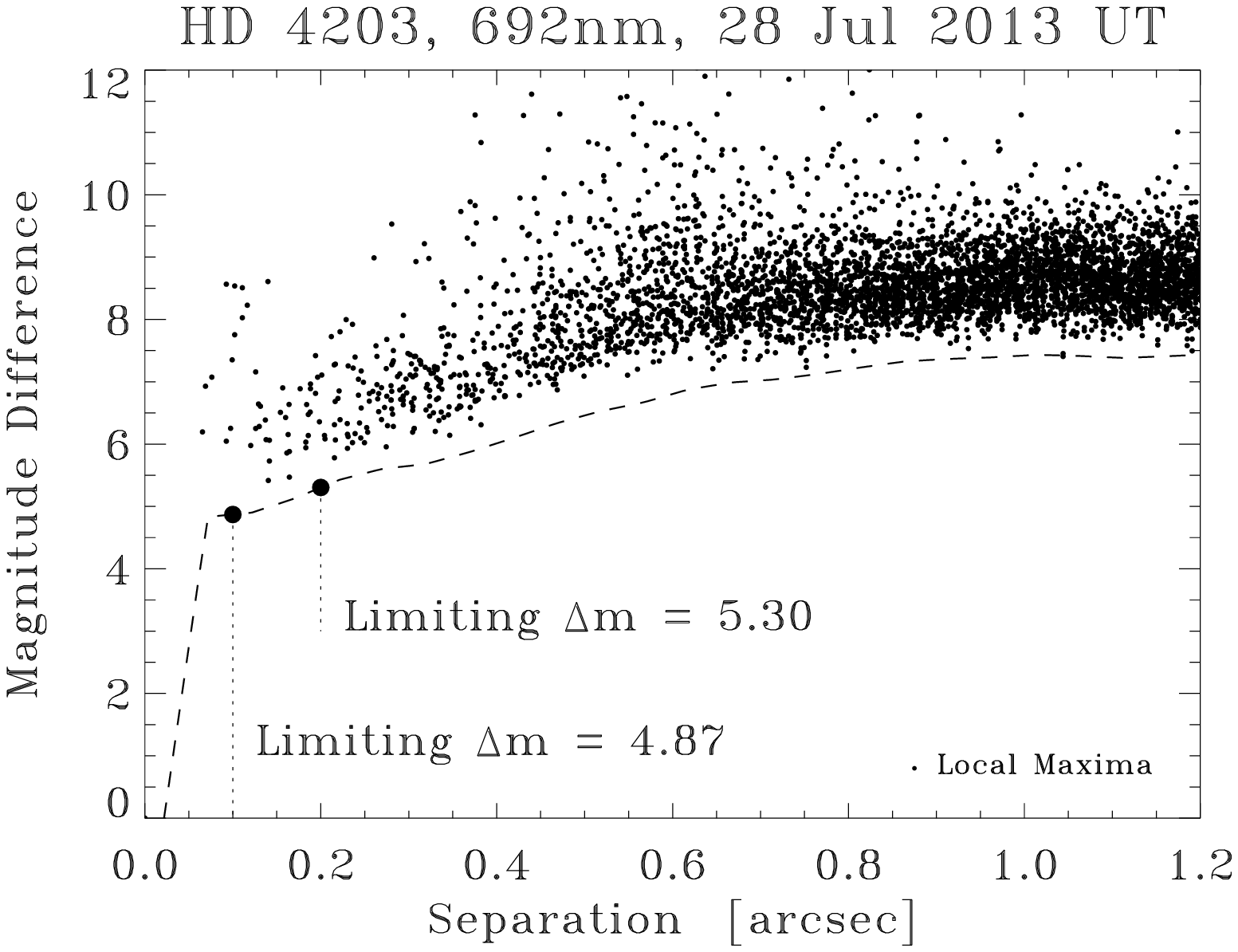} &
      \includegraphics[width=7.2cm]{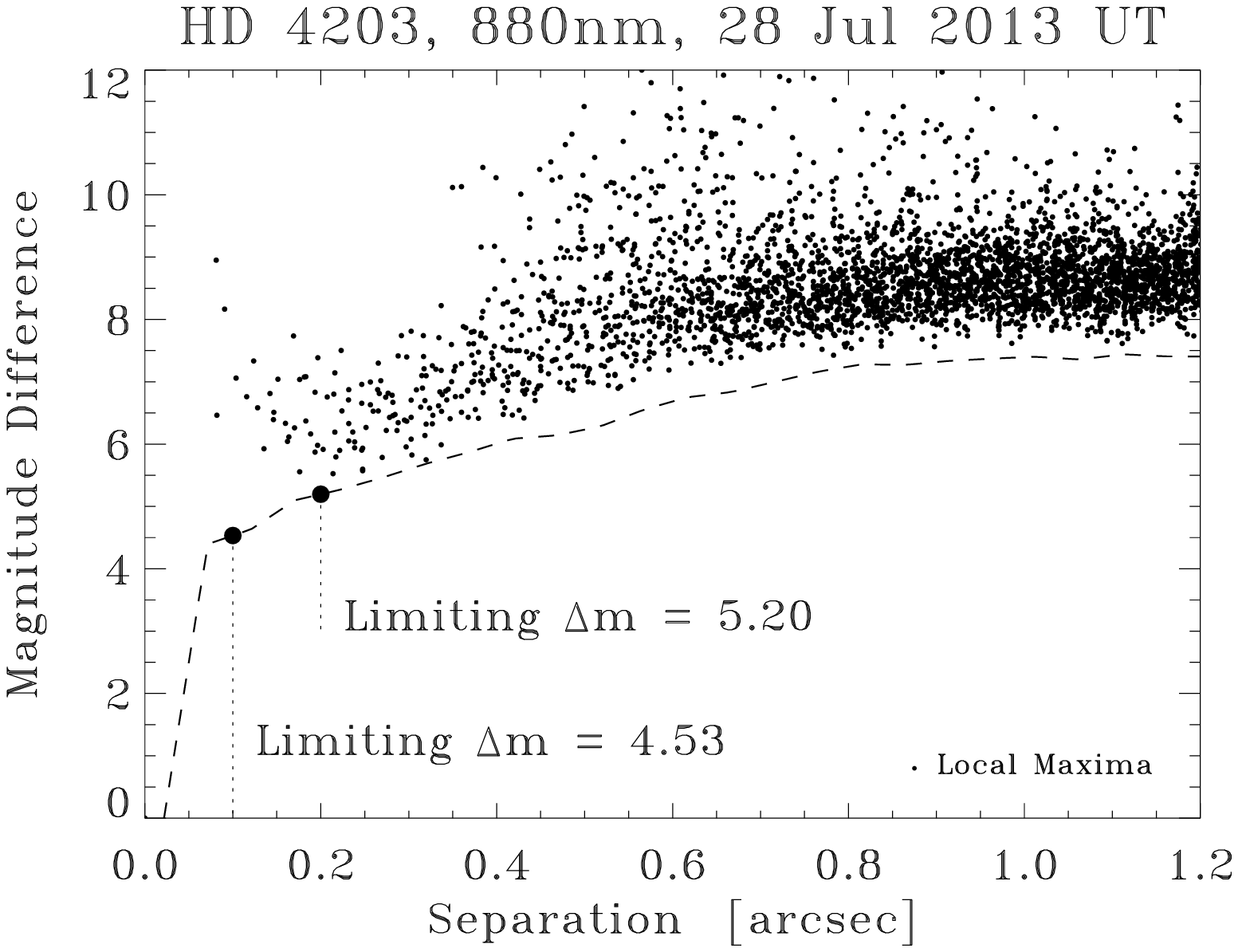} \\
      \includegraphics[width=7.2cm]{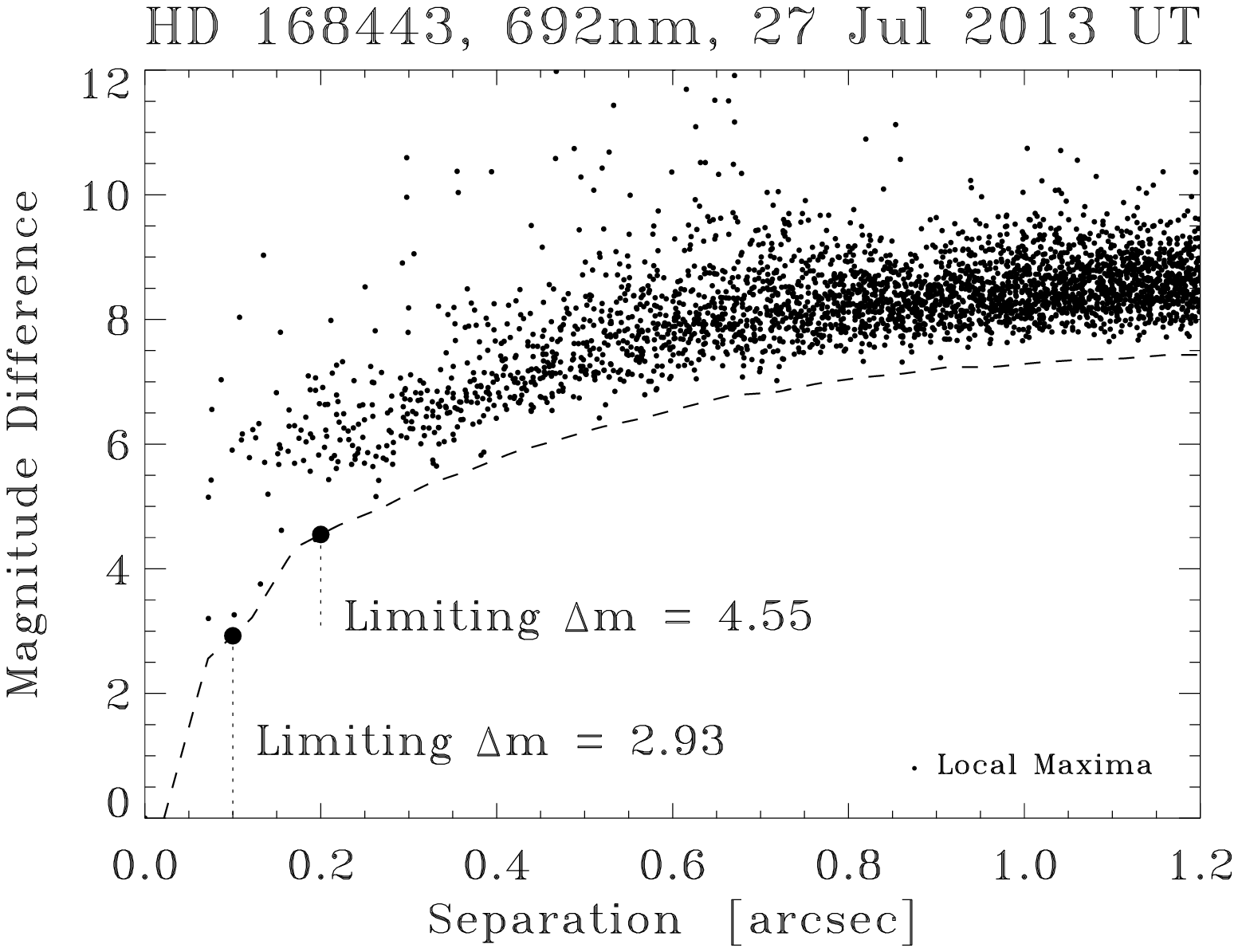} &
      \includegraphics[width=7.2cm]{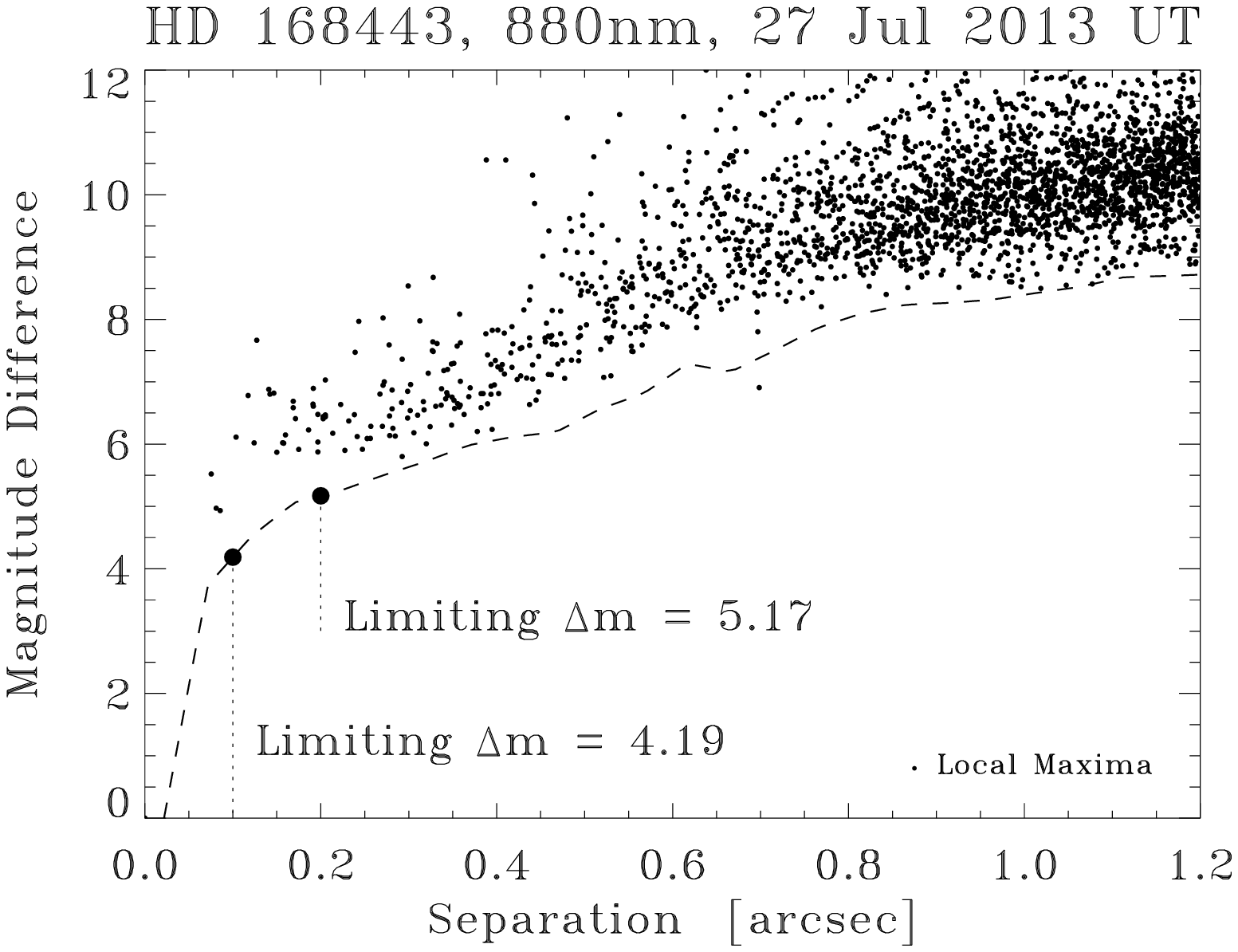} \\
      \includegraphics[width=7.2cm]{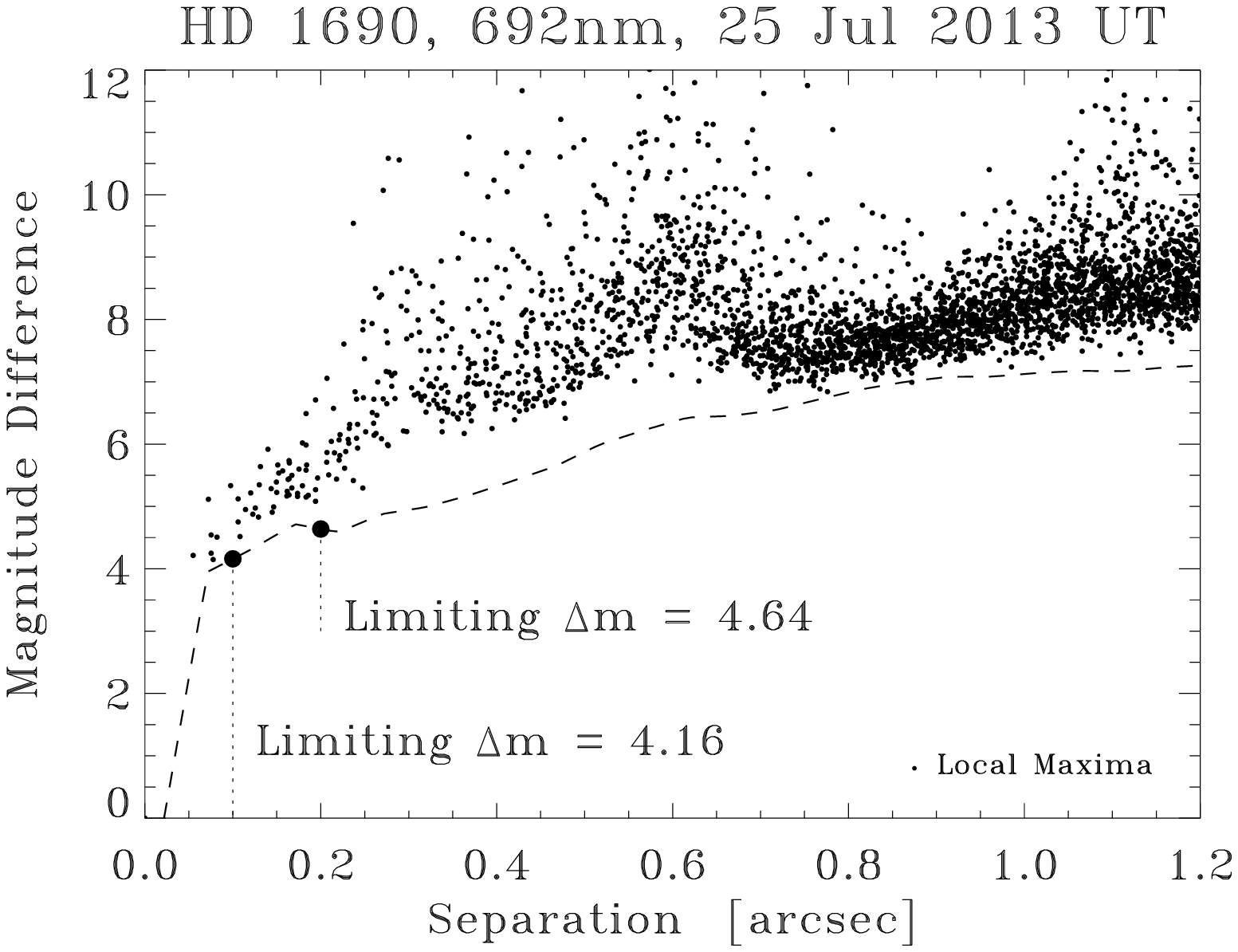} &
      \includegraphics[width=7.2cm]{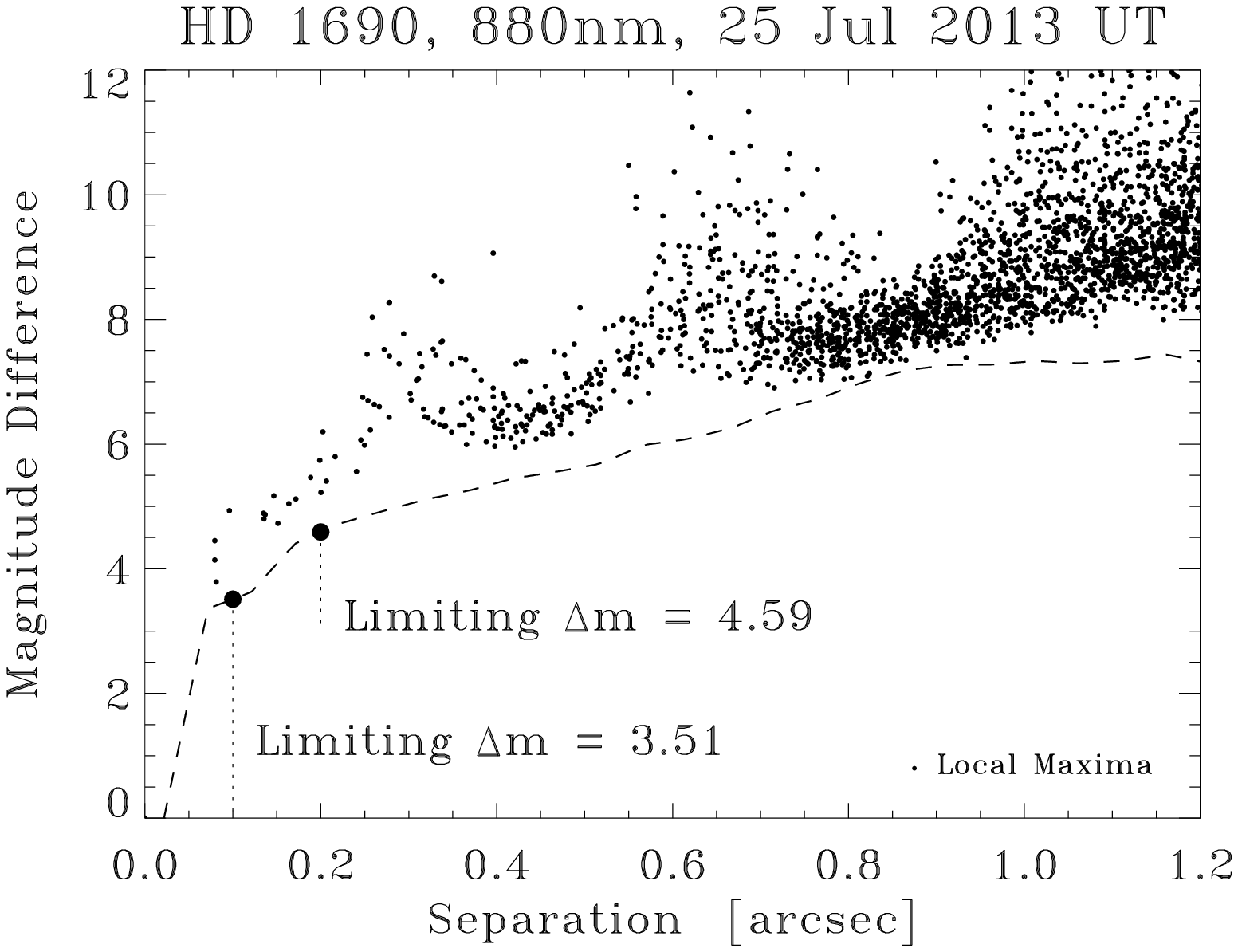} \\
      \includegraphics[width=7.2cm]{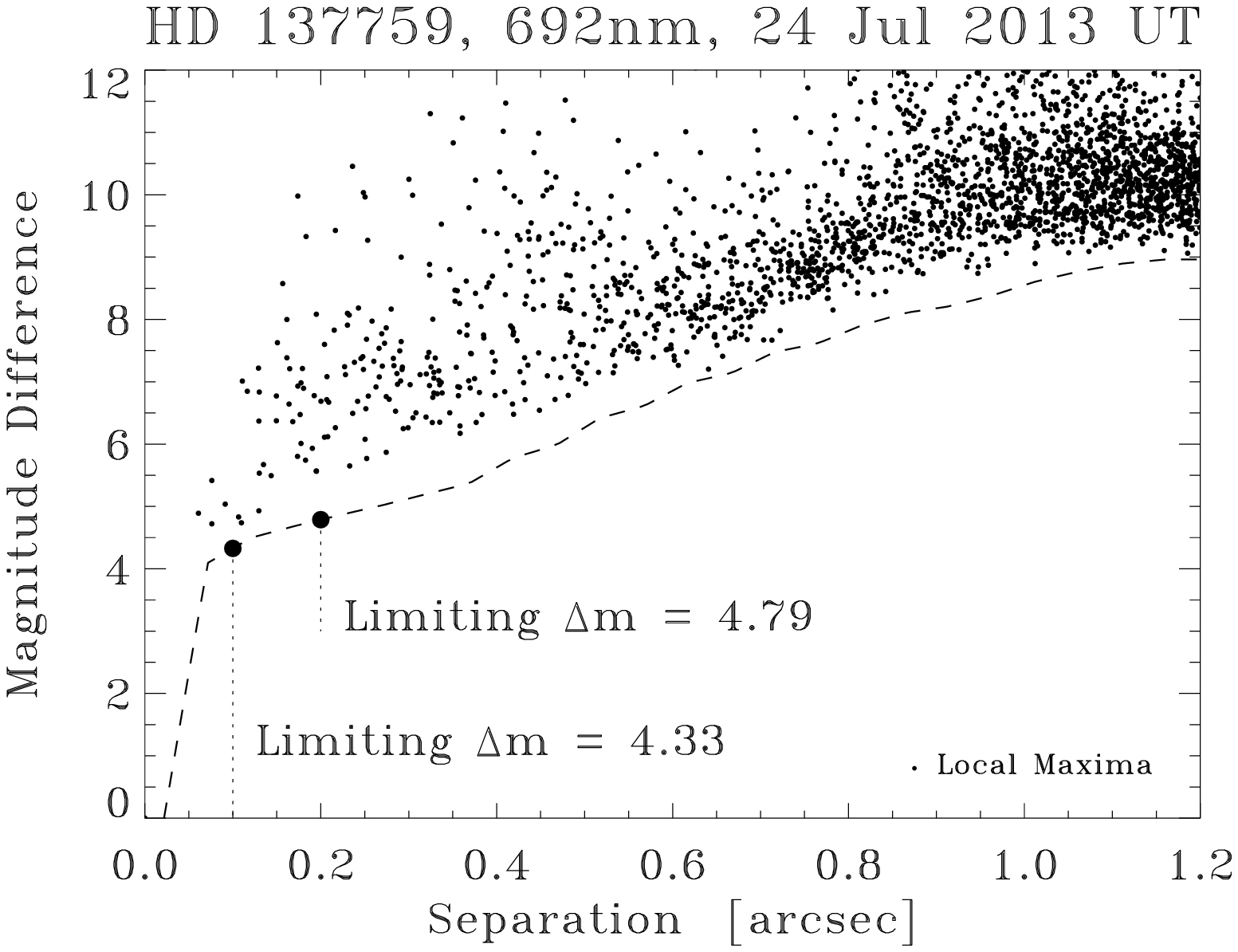} &
      \includegraphics[width=7.2cm]{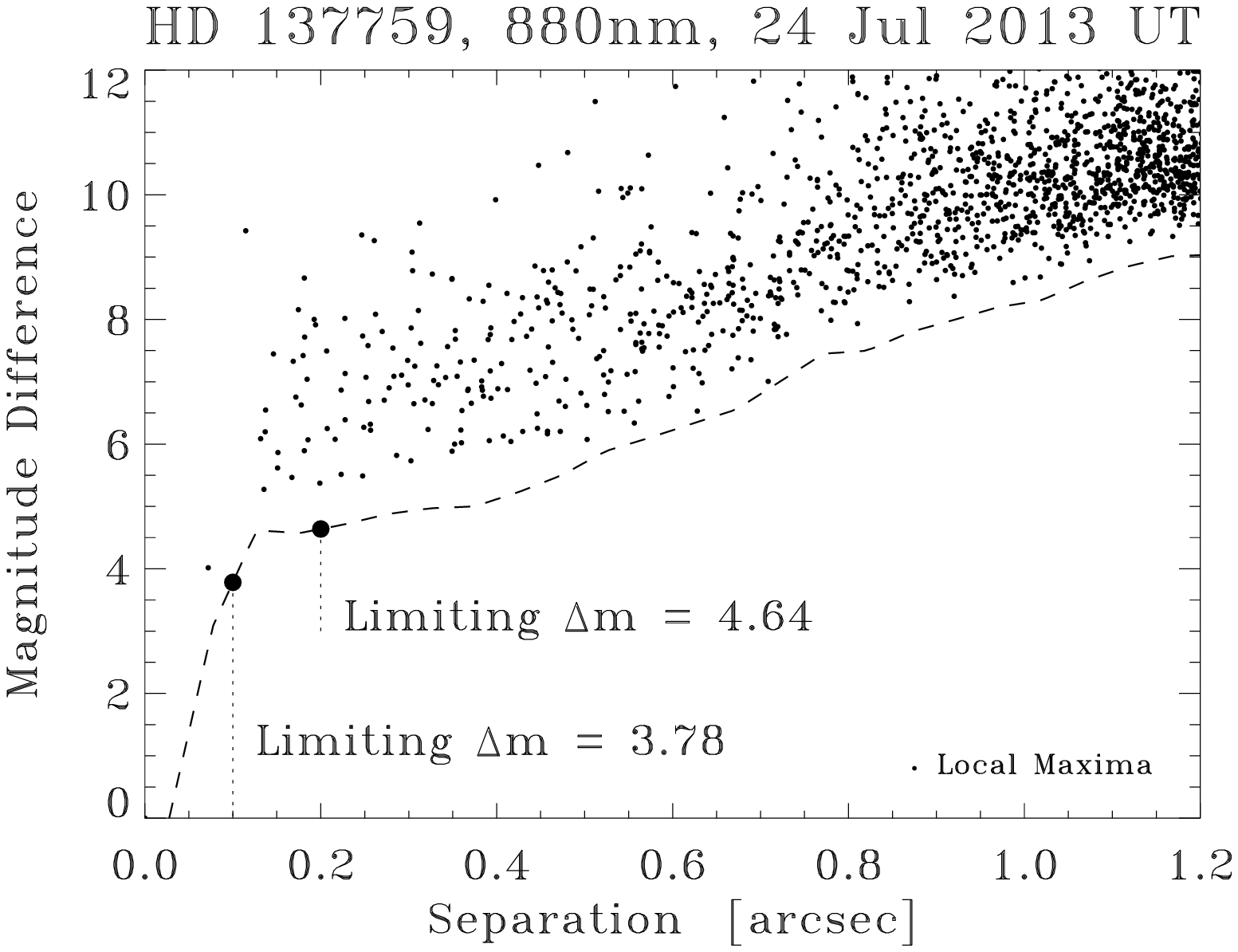}
    \end{tabular}
  \end{center}
  \caption{Limiting magnitude (difference between local maxima and
    minima) as a function of separation from the host star. Each plot
    was calculated from the corresponding image shown in Figure
    \ref{imagesfig}. The dashed line is a cubic spline interpolation
    of the 5$\sigma$ detection limit. Limiting magnitudes are
    explicitly stated for $0.1\arcsec$ and $0.2\arcsec$ in each case.}
  \label{curvesfig}
\end{figure*}

This section presents the results of the imaging observations. The
reduced DSSI speckle images in the two passbands (692~nm and 880~nm)
are shown in Figure \ref{imagesfig}, where the field-of-view is
2.8$\times$2.8$\arcsec$. The limiting magnitude curves based on these
images are shown in Figure \ref{curvesfig}. These show the magnitude
difference between local maxima and minima in the image as a function
of the separation from the central star. Also shown in these plots is
a cubic spline interpolation of the 5$\sigma$ detection limit from
$0.05\arcsec$ to $1.2\arcsec$. Magnitude limits ($\Delta m$) for
stellar companions compared with the magnitude of the primary are
shown for separations of $0.1\arcsec$ and $0.2\arcsec$. These results
are summarized in Table \ref{resultstab}. We describe these separately
for each of the targets and provide a more quantitative analysis of
the detection limits in Section \ref{limits}.

%%%%%%%%%%%%%%%%%%%%%%%%%%%%%%%%%%%%%%%%%%%%%%%%%%%%%%%%%%%%%%%%%%%%

\subsection{HD~4203}

HD~4203 is a $V = 8.7$ magnitude dwarf star at a distance of 77.2~pc
(See Table \ref{paramtab}). The resulting $5\sigma$ detection limits
shown in Table \ref{resultstab} amd Figure \ref{curvesfig} exclude
stellar companions with $\Delta m \sim 4.5$ and $\Delta m \sim 7.5$ at
separations of $0.1\arcsec$ and $1.4\arcsec$ respectively. These
separations correspond to an exclusion range of 3.86--108.1~AU. The
distance modulus for HD~4203 is -4.44. This means we can exclude the
presence of M and L dwarfs at the high separation end.

%M - m = 5 - 5*log10(d)
%      = -4.44
%0.1
%M(692) = -4.44 + 4.87 + 8.7 = 9.13
%M(880) = -4.44 + 4.53 + 8.7 = 8.79
%0.2
%M(692) = -4.44 + 5.30 + 8.7 = 9.56
%M(880) = -4.44 + 5.20 + 8.7 = 9.46

%%%%%%%%%%%%%%%%%%%%%%%%%%%%%%%%%%%%%%%%%%%%%%%%%%%%%%%%%%%%%%%%%%%%

\subsection{HD~168443}

HD~168443 is a $V = 6.92$ magnitude dwarf star and lies 37.4~pc from
the Sun (See Table \ref{paramtab}) with a distance modulus of -2.86.
Based on the DSSI results described in this section, the presence of
stellar companions with with $\Delta m \sim 4.0$ and $\Delta m \sim
8.0$ are excluded at the $5\sigma$ level for separations of
$0.1\arcsec$ and $1.4\arcsec$ respectively. These separations
correspond to an exclusion range of 1.87--52.4~AU. For this target,
the detection limits close to the star ($0.1\arcsec$) are $\sim 50$\%
superior at 880~nm than at 692~nm. Additionally, detection limits at
larger separations reach $\Delta m \sim 9.0$ for 880~nm. As with
HD~4203, this excludes the presence of M and L dwarfs at larger
separations.

Note that there is a single peak which is formally above $5\sigma$
located at $0.7\arcsec$ north of the primary in the 880~nm image but
not detected in the 692~nm image. A similar peak at the same location
in 880~nm images was noticed for other stars during the observing
run. This was caused by a light-leak problem in the 880~nm camera
which resulted in a faint reflection in the 880~nm images. A further
observation of the target will confirm that this is indeed the cause
of the spurious peak.

%M - m = 5 - 5*log10(d)
%      = -2.86
%0.1
%M(692) = -2.86 + 2.93 + 6.92 = 6.99
%M(880) = -2.86 + 4.19 + 6.92 = 8.25
%0.2
%M(692) = -2.86 + 4.55 + 6.92 = 8.61
%M(880) = -2.86 + 5.17 + 6.92 = 9.26

%%%%%%%%%%%%%%%%%%%%%%%%%%%%%%%%%%%%%%%%%%%%%%%%%%%%%%%%%%%%%%%%%%%%

\subsection{HD~1690}

HD~1690 has the weakest exclusion limits of the four targets since is
it a giant star at a larger distance. The star is a $V = 9.17$
magnitude K1 giant at a distance of $\sim 310$~pc (See Table
\ref{paramtab}). The resulting distance modulus is -7.46. Our DSSI
results provide $5 \sigma$ stellar companions exclusion limits of
$\Delta m \sim 4.0$ and $\Delta m \sim 7.5$ for separations of
$0.1\arcsec$ and $1.4\arcsec$ respectively. These separations
correspond to an exclusion range of 15.5--434~AU. This is sufficient
to rule out late M dwarf stellar companions at larger separations.

%M - m = 5 - 5*log10(d)
%      = -7.46
%0.1
%M(692) = -7.46 + 4.16 + 9.17 = 5.87
%M(880) = -7.46 + 3.51 + 9.17 = 5.22
%0.2
%M(692) = -7.46 + 4.64 + 9.17 = 6.35
%M(880) = -7.46 + 4.59 + 9.17 = 6.30

%%%%%%%%%%%%%%%%%%%%%%%%%%%%%%%%%%%%%%%%%%%%%%%%%%%%%%%%%%%%%%%%%%%%

\subsection{HD~137759}

HD~137759 is a bright ($V = 3.29$) K2 giant at a distance of 31~pc
(See Table \ref{paramtab}) with a distance modulus of -2.46. The
$5\sigma$ detection limits resulting from our DSSI observations
exclude stellar companions with $\Delta m \sim 4.0$ and $\Delta m \sim
9.0$ at separations of $0.1\arcsec$ and $1.4\arcsec$
respectively. These separations correspond to an exclusion range of
1.55--43.4~AU. Although this star provided the most favorable
conditions for companion exclusion in terms of proximity to the star
and $\Delta m$, the intrinsic brightness of the host star limits the
range of stars which may realistically be excluded. Based on our
results, we can exclude the presence of M dwarfs at large separations.

%M - m = 5 - 5*log10(d)
%      = -2.46
%0.1
%M(692) = -2.46 + 4.33 + 3.29 = 5.16
%M(880) = -2.46 + 3.78 + 3.29 = 4.61
%0.2
%M(692) = -2.46 + 4.79 + 3.29 = 5.62
%M(880) = -2.46 + 4.64 + 3.29 = 5.47

%%%%%%%%%%%%%%%%%%%%%%%%%%%%%%%%%%%%%%%%%%%%%%%%%%%%%%%%%%%%%%%%%%%%

\section{Limits to Stellar/Planetary Companions}
\label{limits}

The results presented in Section \ref{imagingresults} provide
estimates on the upper limits of stellar companions based on the
speckle images and distance moduli. Here we quantify both the upper
and lower limts using the speckle data results and the RV linear
trends.

The conversion of the speckle $\Delta m$ values to a companion mass
involved the following procedure. Since we know the mass and spectral
type of the primary (see Table \ref{paramtab}), we computed the $V$
magnitude of a possible companion as a function of spectral type and
mass using the mass-radius relationship of \citet{hen93}. The
resulting $\Delta V$ values were then converted to an instrumental
$\Delta m$ in the two speckle filters using the Pickles spectral
library and the DSSI filter transmission curves. The instrumental
$\Delta m$ values for each mass were combined with the detection limit
curves shown in Figure \ref{curvesfig} to generate the limiting mass
as a function of separation.

To estimate the lower limit on stellar or planetary companions to the
target stars, we convert the linear trend in the RV data (shown in
Table \ref{paramtab}) to an acceleration, $\dot{v}$. The law of
gravity can then be used to convert this acceleration into a mass
estimate via $M_p = (\dot{v} a^2) / G$ where we have assumed a
circular orbit for the companion.

\begin{figure*}
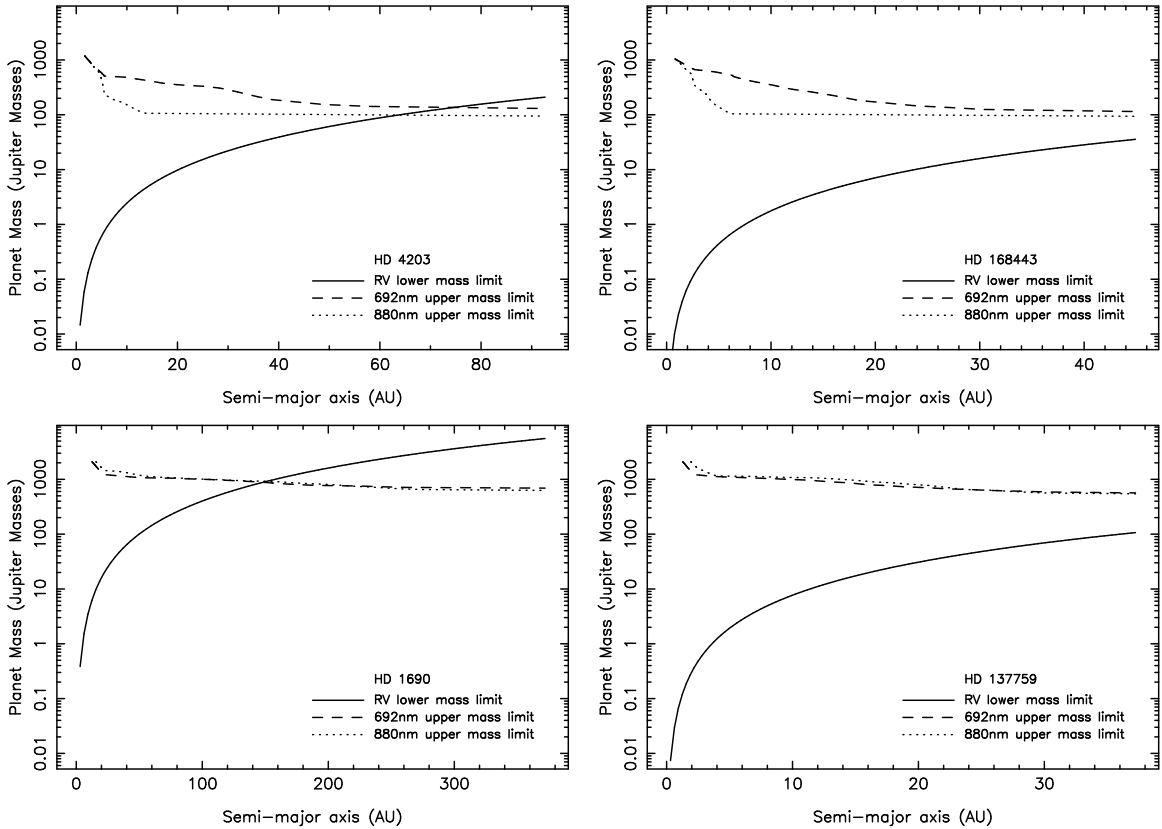

  \begin{center}
    \begin{tabular}{cc}
      \includegraphics[angle=270,width=7.5cm]{f07a.ps} &
      \includegraphics[angle=270,width=7.5cm]{f07b.ps} \\
      \includegraphics[angle=270,width=7.5cm]{f07c.ps} &
      \includegraphics[angle=270,width=7.5cm]{f07d.ps}
    \end{tabular}
  \end{center}
  \caption{Upper and lower limits on the companion mass present in
    each of the four systems as a function of semi-major axis. The
    upper limits are calculated from the imaging results for 692~nm
    (dashed line) and 880~nm (dotted line). The lower limits,
    indicated by the solid lines, are calculated based on the RV
    linear trends.}
  \label{limitsfig}
\end{figure*}

The upper and lower mass limits described above are plotted for each
system in Figure \ref{limitsfig} which show companion mass as a
function of separation. The upper limits derived from the 692~nm and
880~nm observations are represented by the dashed and dotted lines
respectively. The lower limits derived from the RV linear trend are
shown by the solid lines. Anything below the solid curve is either not
massive enough or not close enough to the host star to produce the
measured RV trend. Thus, the region between the speckle curves and the
RV curve are where a companion could exist at this point. Note that at
larger separations, the 880~nm curve is always below the 692~nm curve,
so the loss of speckle correlations in 692~nm for large separations is
not relevant, as the 880~nm data tends to set the upper
limit.

As described in Section \ref{hd4203update}, the linear trend is
adequately explained by the presence of an additional planet. This is
consistent with our null detection of a stellar companion over a broad
range of separations (see Section \ref{imagingresults}). We can thus
be confident that we have indeed detected a second planet in the
HD~4203 system.

The two known planets of the HD~168443 system were discovered by
\citet{mar99,mar01}. The orbital parameters were further refined by
\citet{pil11} along with photometric observations which excluded
transits for the inner planet. The revised parameters of \citet{pil11}
also confirmed a long-term trend in the RV data which indicated the
presence of a long-period companion. The DSSI results shown in Section
\ref{imagingresults} are sensitive out to a separation of 52.4~AU. The
combined upper and lower mass limits shown in Figure \ref{limitsfig}
indicates that any companion within this separation range is highly
likely to be planetary in nature. We thus conclude that the HD~168443
system contains a third planet for which the orbit has yet to be fully
sampled with RV data.

The planet orbiting the giant star HD~1690 was discovered using HARPS
data and published by \citet{mou11}. The DSSI observations exclude
companions over a relatively broad range of separations:
15.5--434~AU. The combined mass limits shown in Figure \ref{limitsfig}
are sufficient to rule out low-mass stars only at large
separations. Additionally, the companion would need to have a
separation smaller than $\sim 20$~AU in order to be of planetary
mass. Thus a stellar companion to the host star could still provide a
viable explanation of the RV trend over a wide range of semi-major
axis values.

The planetary companion to HD~137759 was discovered by \citet{fri02},
and the orbit was further refined by \citet{zec08}, whose radial
velocity data revealed a linear trend in the RV residuals. Further
studies by \citet{kan10} improved the orbital parameters and confirmed
the linear trend as part of the Keplerian orbital solution. Similar to
HD~1690, low-mass stellar companions are ruled out at the separation
limit of the DSSI observations, but still possible at smaller
separations.

%%%%%%%%%%%%%%%%%%%%%%%%%%%%%%%%%%%%%%%%%%%%%%%%%%%%%%%%%%%%%%%%%%%%

\section{A Note on Stellar Abundances}
\label{abundances}

There have been a large number of independent studies, for example
\citet{bon06} and \citet{fis05}, that have consistently confirmed the
enrichment of iron within giant exoplanet host stars. It is still
widely debated whether the apparent increase in stellar iron content
is caused by the exoplanet or whether it is prerequisite for exoplanet
formation. However, the presence of giant exoplanets leaves an
undeniable tracer in the observed stellar iron abundances, where many
have [Fe/H] $>$ 0.3 dex \citep{san04}.

The [Fe/H] content in HD 4203 has been measured by a number of
different groups. The average [Fe/H] value is 0.41 dex, where the
maximum variation between the measured abundances, or {\it spread}, is
0.06 dex (see \citet{hin13}). In other words, HD 4203 displays a
classic enrichment in iron given the presence of a giant exoplanet,
confirmed by multiple groups.

In comparison, HD 168443 has an average [Fe/H] abundance of 0.06 dex,
with a spread in measurements of 0.18 dex between the groups. The
maximum value reported was 0.12 dex by \citet{bru11} while the lowest
was -0.06 dex per \citet{hau05}. HD 137759 had reported similar [Fe/H]
measurements by 4 groups, where the maximum was 0.13 dex
\citep{sad05,ecu06,gil06}, minimum was -0.03 dex \citep{the99}, and
average was [Fe/H] = 0.08 dex. Neither of these stars shows an obvious
iron enrichment, despite the presence of giant exoplanets. Finally,
given the extreme distance of HD 1690, \citet{mou11} was the only
source to report [Fe/H] = -0.32 dex.

Out of the four exoplanet host stars studied here, only HD~4203
exhibits iron enrichment as seen in the majority of exoplanet host
stars. While HD 4203b also has the smallest eccentricity, there does
not appear to be any general correlation between stellar iron content
and exoplanet eccentricity \citep{udr07}.

%%%%%%%%%%%%%%%%%%%%%%%%%%%%%%%%%%%%%%%%%%%%%%%%%%%%%%%%%%%%%%%%%%%%

\section{Conclusions}

The stellar multiplicity of systems which harbor planets is a subject
for which our knowledge base is still evolving. Its relevance plays a
role in many aspects of planetary formation and evolution. Here we
have presented the results of an imaging investigation of four
specific systems where at least one of the planets is of high
eccentricity. In all four cases we have been able to rule out the
presence of all but very low-mass stars over a wide range of
separations. For HD~4203, the DSSI data in combination with Keck/HIRES
data confirm that there is indeed a second planet harbored in that
system. We can therefore state with confidence that HD~4203 no longer
has a detected RV trend without explanation. For HD~168443, the extent
of the RV linear trend and the limits placed by the DSSI results
strongly imply that a third planet is the best explanation for the
observed trend. HD~1690 and HD~137759 are both giant stars and so the
detection limits are not quite so deep as they are for dwarfs. The
DSSI results thus rule out solar-mass stars as companions to these
giant hosts but leave ambiguity as to whether a massive planet or
brown dwarf at a variety of separations could explain the linear
trend, although low-mass stellar companions are ruled out at large
separations.

The lack of detected stellar companions in all cases returns one to
the original question of determining the source of the high planetary
eccentricities. The presence of a wide-binary companion has long been
considered as a possible source of orbital perturbation
\citep{zak04,mal07,kai13}. The observations presented in this work do
not exclude the possibility of a stellar companion at even wider
separations that may have been the catalyst for past dynamical
interactions within the system. For example, the 2MASS All-Sky Point
Source Catalog \citep{skr06} does not reveal any sources within
$10\arcsec$ of each target. Our results also do not take into account
a passing star that may have induced a similar transfer of angular
momentum. A more common approach to explaining the diversity of
exoplanetary eccentricities is the occurrence of past planet-planet
scattering events \citep{ada03,cha08,jur08}. Given that we detect no
stellar companions for our four systems, this appears to be consistent
with ``internal'' planetary interactions.

Finally, we have shown the utility of instruments such as DSSI for
conducting studies of the brightest exoplanet host stars to determine
multiplicity. Continued observations will prove essential to
understanding the multiplicity of exoplanet host stars which are
closest to the Sun and thus place the formation our Solar System and
the singularity of our own star in context.

%%%%%%%%%%%%%%%%%%%%%%%%%%%%%%%%%%%%%%%%%%%%%%%%%%%%%%%%%%%%%%%%%%%%

\section*{Acknowledgements}

Based on observations obtained at the Gemini Observatory, which is
operated by the Association of Universities for Research in Astronomy,
Inc., under a cooperative agreement with the NSF on behalf of the
Gemini partnership: the National Science Foundation (United States),
the National Research Council (Canada), CONICYT (Chile), the
Australian Research Council (Australia), Minist\'{e}rio da
Ci\^{e}ncia, Tecnologia e Inova\c{c}\~{a}o (Brazil) and Ministerio de
Ciencia, Tecnolog\'{i}a e Innovaci\'{o}n Productiva (Argentina). This
research has made use of the Exoplanet Orbit Database and the
Exoplanet Data Explorer at exoplanets.org. This research has also made
use of the NASA/ IPAC Infrared Science Archive, which is operated by
the Jet Propulsion Laboratory, California Institute of Technology,
under contract with the National Aeronautics and Space Administration.
The authors acknowledge financial support from the National Science
Foundation through grant AST-1109662.

%%%%%%%%%%%%%%%%%%%%%%%%%%%%%%%%%%%%%%%%%%%%%%%%%%%%%%%%%%%%%%%%%%%%


\begin{thebibliography}{}

\bibitem[\protect\citeauthoryear{Abt \& Levy}{1976}]{abt76} Abt, H.A.,
  Levy, S.G. 1976, ApJS, 30, 273
\bibitem[\protect\citeauthoryear{Adams \& Laughlin}{2003}]{ada03}
  Adams, F.C., Laughlin, G. 2003, Icarus, 163, 290
\bibitem[\protect\citeauthoryear{Bergfors et al.}{2013}]{ber13}
  Bergfors, C., et al. 2013, MNRAS, 428, 182
\bibitem[\protect\citeauthoryear{Bond et al.}{2006}]{bon06} Bond,
  J.C., Tinney, C.G., Butler, R.P., Jones, H.R.A., Marcy, G.W., Penny,
  A.J., Carter, B.D. 2006, MNRAS, 370, 163
\bibitem[\protect\citeauthoryear{Brugamyer et al.}{2011}]{bru11}
  Brugamyer, E., Dodson-Robinson, S.E., Cochran, W.D., Sneden,
  C. 2011, ApJ, 738, 97
\bibitem[\protect\citeauthoryear{Butler et al.}{2006}]{but06} Butler,
  R.P., et al. 2006, ApJ, 646, 505
\bibitem[\protect\citeauthoryear{Cassan et al.}{2012}]{cas12} Cassan,
  A., et al. 2012, Nature, 481, 167
\bibitem[\protect\citeauthoryear{Chatterjee et al.}{2008}]{cha08}
  Chatterjee, S., Ford, E.B., Matsumura, S., Rasio, F.A. 2008, ApJ,
  686, 580
\bibitem[\protect\citeauthoryear{Crepp et al.}{2012}]{cre12} Crepp,
  J.R., et al. 2012, ApJ, 761, 39
\bibitem[\protect\citeauthoryear{Crepp et al.}{2013}]{cre13} Crepp,
  J.R., Johnson, J.A., Howard, A.W., Marcy, G.W., Gianninas, A.,
  Kilic, M., Wright, J.T. 2013, ApJ, 774, 1
\bibitem[\protect\citeauthoryear{Dressing \&
    Charbonneau}{2013}]{dre13} Dressing, C.D., Charbonneau, D. 2013,
  ApJ, 767, 95
\bibitem[\protect\citeauthoryear{Duquennoy \& Mayor}{1991}]{duq91}
  Duquennoy, A., Mayor, M. 1991, A\&A, 248, 485
\bibitem[\protect\citeauthoryear{Ecuvillon et al.}{2006}]{ecu06}
  Ecuvillon, A., Israelian, G., Santos, N.C., Mayor, M., Gilli,
  G. 2006, A\&A, 449, 809
\bibitem[\protect\citeauthoryear{Eggenberger et al.}{2004}]{egg04}
  Eggenberger, A., Udry, S., Mayor, M. 2004, A\&A, 417, 353
\bibitem[\protect\citeauthoryear{Eggenberger et al.}{2007}]{egg07}
  Eggenberger, A., Udry, S., Chauvin, G., Beuzit, J.-L., Lagrange,
  A.-M., S\'egransan, D., Mayor, M. 2007, A\&A, 474, 273
\bibitem[\protect\citeauthoryear{Fischer \& Valenti}{2005}]{fis05}
  Fischer, D.A., Valenti, J. 2005, ApJ, 622, 1102
\bibitem[\protect\citeauthoryear{Frink et al.}{2002}]{fri02} Frink,
  S., Mitchell, D.S., Quirrenbach, A., Fischer, D.A., Marcy, G.W.,
  Butler, R.P. 2002, ApJ, 576, 478
\bibitem[\protect\citeauthoryear{Gilli et al.}{2006}]{gil06} Gilli,
  G., Israelian, G., Ecuvillon, A., Santos, N.C., Mayor, M. 2006,
  A\&A, 449, 723
\bibitem[\protect\citeauthoryear{Henry \& McCarthy}{1993}]{hen93}
  Henry, T.J., McCarthy, D.W., Jr. 1993, AJ, 106, 773
\bibitem[\protect\citeauthoryear{Hinkel \& Kane}{2013}]{hin13} Hinkel,
  N.R., Kane, S.R. 2013, MNRAS, 432, L36
\bibitem[\protect\citeauthoryear{Horch et al.}{2009}]{hor09} Horch,
  E.P., Veillette, D.R., Baena Galle, R., Shah, S.C, O'Rielly, G.V.,
  van Altena, W.F. 2009, AJ, 137, 5057
\bibitem[\protect\citeauthoryear{Horch et al.}{2011}]{hor11} Horch,
  E.P., Gomez, S.C., Sherry, W.H., Howell, S.B., Ciardi, D.R.,
  Anderson, L.M., van Altena, W.F. 2011, AJ, 141, 45
\bibitem[\protect\citeauthoryear{Horch et al.}{2012}]{hor12} Horch,
  E.P., Howell, S.B., Everett, M.E., Ciardi, D.R. 2012, AJ, 144, 165
\bibitem[\protect\citeauthoryear{Howell et al.}{2011}]{how11} Howell,
  S.B., Everett, M.E., Sherry, W., Horch, E., Ciardi, D.R. 2011, AJ,
  142, 19
\bibitem[\protect\citeauthoryear{Huang et al.}{2005}]{hau05} Huang,
  C., Zhao, G., Zhang, H.W., Chen, Y.Q. 2005, MNRAS, 363, 71
\bibitem[\protect\citeauthoryear{Juri\'c \& Tremaine}{2008}]{jur08}
  Juri\'c, M., Tremaine, S., 2008, ApJ, 686, 603
\bibitem[\protect\citeauthoryear{Kaib et al.}{2013}]{kai13} Kaib,
  N.A., Raymond, S.N., Duncan, M. 2013, Nature, 493, 381
\bibitem[\protect\citeauthoryear{Kane et al.}{2010}]{kan10} Kane,
  S.R., Reffert, S., Henry, G.W., Fischer, D., Schwab, C., Clubb,
  K.I. 2010, ApJ, 720, 1644
\bibitem[\protect\citeauthoryear{Lohmann et al.}{1983}]{loh83}
  Lohmann, A.W., Weigelt, G., Wirnitzer, B. 1983, ApOpt, 22, 4028
\bibitem[\protect\citeauthoryear{Malmberg et al.}{2007}]{mal07}
  Malmberg, D., de Angeli, F., Davies, M.B., Church, R.P., Mackey, D.,
  Wilkinson, M.I. 2007, MNRAS, 378, 1207
\bibitem[\protect\citeauthoryear{Marcy et al.}{1999}]{mar99} Marcy,
  G.W., Butler, R.P., Vogt, S.S., Fischer, D., Liu, M.C. 1999, ApJ,
  520, 239
\bibitem[\protect\citeauthoryear{Marcy et al.}{2001}]{mar01} Marcy,
  G.W., et al. 2001, ApJ, 555, 418
\bibitem[\protect\citeauthoryear{Meng et al.}{1990}]{men90} Meng, J.,
  Aitken, G.J.M., Hege, E.K., Morgan, J.S. 1990, JOSAA, 7, 1243
\bibitem[\protect\citeauthoryear{Moutou et al.}{2011}]{mou11} Moutou,
  C., et al. 2011, A\&A, 527, 63
\bibitem[\protect\citeauthoryear{Pilyavsky et al.}{2011}]{pil11}
  Pilyavsky, G., et al. 2011, ApJ, 743, 162
\bibitem[\protect\citeauthoryear{Press et al.}{2002}]{pre02}
  Press, W.H., Teukolsky, S.A., Vetterling, W.T., Flannery,
  B.P. 2002, Numerical recipes in C++ : the art of scientific
  computing by William H. Press. xxviii, 1,002 p.~: ill.~; 26 cm.~
  Includes bibliographical references and index.~ISBN : 0521750334
\bibitem[\protect\citeauthoryear{Raghavan et al.}{2010}]{rag10}
  Raghavan, D., et al. 2010, ApJS, 190, 1
\bibitem[\protect\citeauthoryear{Roell et al.}{2012}]{roe12} Roell,
  T., Neuh\"auser, R., Seifahrt, A., Mugrauer, M. 2012, A\&A, 542, 92
\bibitem[\protect\citeauthoryear{Sadakane et al.}{2005}]{sad05}
  Sadakane, K., Ohnishi, T., Ohkubo, M., Takeda, Y. 2005, PASJ, 57,
  127
\bibitem[\protect\citeauthoryear{Santos et al.}{2004}]{san04} Santos,
  N.C., Israelian, G., Mayor, M. 2004, A\&A, 415, 1153
\bibitem[\protect\citeauthoryear{Skrutskie et al.}{2006}]{skr06}
  Skrutskie, M.F., et al. 2006, AJ, 131, 1163
\bibitem[\protect\citeauthoryear{Th\'evenin \& Idiart}{1999}]{the99}
  Th\'evenin, F., Idiart, T.P. 1999, ApJ, 521, 753
\bibitem[\protect\citeauthoryear{Udry \& Santos}{2007}]{udr07} Udry,
  S., Santos, N.C. 2007, ARA\&A, 45, 397
\bibitem[\protect\citeauthoryear{van Leeuwen}{2007}]{van07} van
  Leeuwen, F. (ed.) 2007, Hipparcos, the New Reduction of the Raw Data
  (Astrophysics and Space Science Library, Vol. 350)
\bibitem[\protect\citeauthoryear{Vogt et al.}{1994}]{vog94} Vogt,
  S.S., et al. 1994, \procspie, 2198, 362
\bibitem[\protect\citeauthoryear{Vogt et al.}{2002}]{vog02} Vogt,
  S.S., Butler, R.P., Marcy, G.W., Fischer, D.A., Pourbaix, D., Apps,
  K., Laughlin, G. 2002, ApJ, 568, 352
\bibitem[\protect\citeauthoryear{Wang et al.}{2012}]{wan12} Wang,
  S.X., et al. 2012, ApJ, 761, 46
\bibitem[\protect\citeauthoryear{Wright \& Howard}{2009}]{wri09}
  Wright, J.T., Howard, A.W. 2009, ApJS, 182, 205
\bibitem[\protect\citeauthoryear{Wright et al.}{2011}]{wri11}
  Wright, J.T., et al. 2011, PASP, 123, 412
\bibitem[\protect\citeauthoryear{Zakamska \& Tremaine}{2004}]{zak04}
  Zakamska, N.L., Tremaine, S. 2004, AJ, 128, 869
\bibitem[\protect\citeauthoryear{Zechmeister et al.}{2008}]{zec08}
  Zechmeister, M., Reffert, S., Hatzes, A.P., Endl, M., Quirrenbach,
  A. 2008, A\&A, 491, 531
\bibitem[\protect\citeauthoryear{Zucker \& Mazeh}{2002}]{zuc02}
  Zucker, S., Mazeh, T. 2002, ApJ, 568, L113

\end{thebibliography}
\end{document}